\documentclass[10pt]{article}
\usepackage[super,compress]{cite}
\usepackage{graphicx}
\usepackage{multirow}
\usepackage{slashed}
\usepackage{amsmath}
\usepackage{amssymb}
\usepackage{titlesec}

\titleformat{\section}
  {\bf}{\thesection}{1em}{}
\titleformat{\subsection}
  {\bfseries \itshape}{\thesubsection}{1em}{}
\titleformat{\subsubsection}
  {\it}{\thesubsubsection}{1em}{}

\newcommand{\pref}[1]{(\ref{#1})}
\newcommand{\mf}{\langle x\rangle_{u-d}}
\newcommand{\hm}{\langle x \rangle_{\Delta u - \Delta d}}
\newcommand{\tm}{\langle x \rangle_{\delta u - \delta d}}

\newcommand{\mpi}{M_{\pi}}
\newcommand{\mN}{M_N}

\newcommand{\OV}[1]{V^{a}_{{#1}}}
\newcommand{\OA}[1]{A^a_{{#1}}}
\newcommand{\OT}[1]{T^a_{{#1}}}
\newcommand{\tq}{{\tilde{q}}}
\newcommand{\gA}{g_A}

\newcommand{\gAbn}{\bar{g}_{A,n}}

\begin{document}
\markboth{Oliver B{\"a}r}{Nucleon-pion state-contamination in lattice QCD}

%
%

\title{ \normalsize \bf Chiral perturbation theory and nucleon-pion-state contaminations in lattice QCD}

\author{\normalsize Oliver B{\"a}r\\[0.2ex]
\it \small Institut f{\"u}r Physik, Humboldt Universit{\"a}t zu Berlin,\\
\it\small Newtonstra{\ss}e 15, 12489 Berlin, Germany\\
\it \small obaer@physik.hu-berlin.de
}
\date{}

\maketitle

\begin{abstract}
{\small
Multi-particle states with additional pions are expected to be a non-negligi\-ble source of excited-state contamination in lattice simulations at the physical point. It is shown that baryon chiral perturbation theory  can be employed to calculate the contamination due to two-particle nucleon-pion states in various nucleon observables. Leading order results are presented for the nucleon axial, tensor and scalar charge and three Mellin moments of parton distribution functions (quark momentum fraction, helicity and transversity moment). Taking into account phenomenological results for the charges and moments the impact of the nucleon-pion-states on lattice estimates for these observables can be estimated. The nucleon-pion-state contribution results in an overestimation of all charges and moments obtained with the plateau method. The overestimation is at the 5-10\% level for source-sink separations of about 2 fm. The source-sink separations accessible in contemporary lattice simulations are found to be too small for chiral perturbation theory to be directly applicable.    
}
\end{abstract}




\section{Introduction}	

Chiral perturbation theory (ChPT) has been a valuable tool in the analysis of Lattice Quantum Chromodynamics (QCD) data for many years. Prime examples are the calculation of finite-volume (FV) effects or the light quark mass dependence of physical observables, to name just the two most prominent applications. ChPT results concerning the quark mass dependence have been commonly used in the chiral extrapolation to relate unphysical lattice results obtained at heavy quark masses to the so-called physical point with quark masses as light as in Nature. However, the need for a chiral extrapolation has been slowly fading away. Constant progress in computer power as well as advances in simulation algorithms have made lattice simulations possible with light quark masses set to their physical values. Such {\em physical point  simulations} do not need a chiral extrapolation and eliminate one major source of systematic error. This benefit is certainly worth the high numerical prize one still has to pay for this kind of simulations.

In this review I report on a different application of ChPT to Lattice QCD, namely the study of excited-state contaminations. With the up and down quark masses  close to their physical values one can expect multi-particle states with additional pions to become a non-negligible excited-state contamination in many correlation functions measured on the lattice to obtain physical observables. The calculation of nucleon observables is a familiar example where excited-state contaminations are known to be a source of significant systematic uncertainty. The multi-particle states expected to be most relevant in this case are two-particle nucleon-pion ($N\pi$) and three-particle $N\pi\pi$ states, and their contribution is accessible within ChPT.

The idea for this application is not new. In Refs.\ \citen{Bar:2012ce} the three-pion-state contribution to the  2-point (2pt)  functions of the axial-vector and pseudo-scalar quark bilinears were computed in ChPT. Probably the first account of this idea is the computation of the $N\pi$-state contamination to the nucleon axial charge in Ref.\ \citen{Tiburzi:2009zp}. This was an attempt to explain, at least qualitatively, why most lattice results at that time underestimated the experimental value of the nucleon axial charge. However, it turned out that the $N\pi$ state contamination leads to an overestimation of the axial charge, and this might be the reason why Ref.\ \citen{Tiburzi:2009zp} remained fairly unnoticed. 

Today results for other nucleon charges and for some first moments of parton distribution functions are available.\cite{Bar:2015zwa,Tiburzi:2015tta,Bar:2016uoj,Bar:2016jof} These results improve on some of the approximations made in the first calculation. The discreteness of the nucleon and pion momenta due to a finite spatial volume is taken into account. In addition, the mapping of smeared interpolating fields, commonly used in lattice calculations, to ChPT has been put on firmer grounds.\cite{Luscher:2013vga,Bar:2013ora}. The goal of this review is to summarize the results of Refs.\ \citen{Bar:2015zwa,Bar:2016uoj,Bar:2016jof}. Some of these results were already presented at workshops and conferences.\cite{Bar:BetheForum,Bar:2015zha, Bar:Niigata} Answers to some of the questions asked at these occasions are also given here. 

\section{$N\pi$-state contributions in nucleon correlation function}

To start with let us consider the nucleon 2pt function, 
 \begin{eqnarray}
 \label{Def2ptfunc}
G_{\rm 2pt}(t)& =& \int_{L^3} {\rm d}^3{{x}}\, \langle  N(\vec{x},t) \overline{N}(0,0)\rangle\,,
\end{eqnarray}
that is  measured in Lattice QCD to compute the nucleon mass.\footnote{We use a continuum notation in this review even if we explicitly refer to lattice correlators.}
Here $N,\overline{N}$ denote interpolating fields with the quantum numbers of the nucleon. The integration over the finite spatial volume $V=L^3$ serves projects to zero spatial momentum. 
Performing the standard spectral decomposition the 2pt function is given as a sum of exponentials,
\begin{equation}
\label{ExpAnsatz}
G_{\rm 2pt}(t) = c_0 e^{-E_0 t} + c_1 e^{-E_1 t} + c_2 e^{-E_2 t} + \ldots\,
\end{equation}
where the ordering $E_0 < E_1 < E_2 < \ldots $ for the energies is assumed. 
The first exponential provides, by construction, the exponential decay with the nucleon mass, $E_0=M_N$. The coefficient $c_0$ is the squared matrix element of the interpolating field between the vacuum and the nucleon at rest.
All the other terms stem from excited states with the same quantum numbers as the nucleon, and the coefficients $c_j$  involve these excited states instead of the nucleon state.

In lattice simulations one usually computes the effective nucleon mass, defined as the negative time derivative of $\ln G_{\rm 2pt}(t)$ .  With \pref{ExpAnsatz} we obtain ($\Delta E_k = E_k-M_N$)
\begin{eqnarray}
\label{Meff}
M_{N,{\rm eff}} & =&  M_N + \frac{c_1}{c_0} \Delta E_1 e^{-\Delta E_{1} t} + \frac{c_2}{c_0} \Delta E_2 e^{-\Delta E_{2} t}+\ldots \,.
\end{eqnarray}
Sending $t$ to infinity the effective mass converges to a constant, the nucleon mass. For finite $t$ the excited-state contribution is exponentially suppressed. Still,  the euclidean time separation needs to be sufficiently large for the excited-state contamination in the correlation function to be small. However, the signal-to-noise problem prevents one to go to very large euclidean times $t$ to make the excited-state contribution arbitrarily small. The smaller the pion mass the smaller is the time separation for which the effective mass can be measured with small statistical errors.\cite{Parisi:1983ae,Lepage:1989hd}  In practice one is so far limited to time separations of about 1 to 1.5 fm, and at these distances the excited-state contribution is still visible in the lattice data.

\begin{figure}[tb]
\centerline{\includegraphics[scale=0.4]{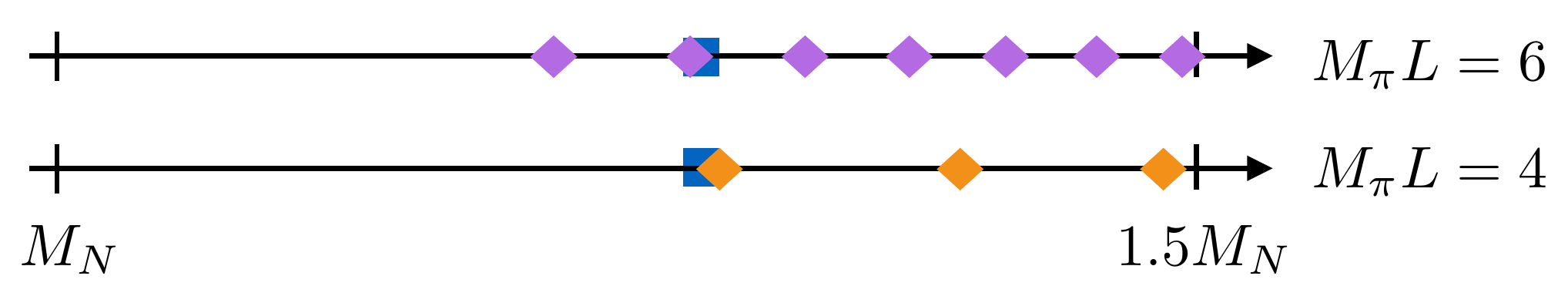}}
\caption{Sketch of the energies of the lowest $N\pi$ states (diamonds) that contribute to the nucleon 2pt function. Discrete momenta corresponding to two $M_{\pi}L$ values are considered and interaction energies are ignored. The blue square represents the energy of the $N\pi\pi$ state with all three particles at rest. }
\label{fig:sketch}
\end{figure}

It is useful to distinguish the excited states that can contribute in a finite spatial volume to the 2pt function. There is a contribution from resonance states that are associated to the nucleon resonances in infinite volume, the most prominent one being the Roper resonance $N^*(1440)$.\footnote{How to determine infinite volume resonance properties from the finite volume energy spectrum is shown in Ref.\ \citen{Luscher:1991cf}.} In addition there is the contribution of multi-particle states, e.g.\ 2-particle $N\pi$ states, 3-particle $N\pi\pi$ states etc. In particular the multi-particle-state contribution is expected to become important for physical pion masses. This is easily seen by ignoring the interaction energy of the particles, which is expected to be rather small since the pions interact only weakly with the nucleon and with themselves. Within this approximation the energy of the $N\pi\pi$ state with all three particles at rest is equal to $M_N+2M_{\pi}\approx 1.3 M_N$. Two-particle $N\pi$ states also contribute to the excited state contamination. Because of parity both nucleon and pion cannot be at rest but need non-vanishing and opposite spatial momenta. If we assume periodic boundary conditions for the finite spatial volume the spatial momenta are discrete, $\vec{p}_k=2\pi\vec{k}/L$, with $\vec{k}$ having having integer-valued components. The larger the lattice extent $L$ the smaller are the discrete energies of the $N\pi$ states allowed by the periodic boundary conditions. If we assume the typical value $M_{\pi}L=4$ we find three $N\pi$ states with energy less than $1.5M_N$. The PACS collaboration carries out lattice simulations with almost physical pion mass satisfying $M_{\pi}L=6$. In that case seven $N\pi$ states have an energy less than $1.5M_N$ (see fig.\ \ref{fig:sketch}). The main conclusion is that quite a few multi-particle states are expected to contribute to the sum in eq.\ (\ref{ExpAnsatz}) before the first resonance state appears. Thus, the multi-particle states dominate the excited-state contribution in the asymptotic regime of large but still finite $t$. 

The impact of the multi-particle states to the 2pt function depends also on the size of the coefficients $c_j, j\ge1$. Not much is  a priori known about these coefficients, except for the fact that the coefficient associated with an $n$-particle state is, in a finite spatial volume, suppressed by $(1/L^3)^{n-1}$. This volume-suppression is sometimes taken as an argument in favor of small multi-particle-state contributions in case of large $L$. However, this argument is not fully convincing. Although the contribution of each individual multi-particle state diminishes, more and more states contribute the larger the volume gets. After all, the multi-particle state contribution is not a finite volume (FV) effect that will disappear in the infinite-volume limit.  

In practice one often tries to account for the excited-state contributions by using multi-state fit ansaetze to analyze lattice data. Recently, up to three excited states were kept in \pref{ExpAnsatz} for the analysis of the 2pt function.\cite{Yoon:2016jzj} However, in fits with more than one excited state the coefficients and mass gaps are very often found to be ill-determined. Although in these cases the fits can be stabilized with a Bayesian analysis and priors for the coefficients and mass gaps, multi-state fits including a few excited states are in general difficult to carry out. 

As already mentioned in the introduction, ChPT can be employed to obtain estimates for the coefficients associated with multi-particle states involving the nucleon and additional pions. This by itself is perhaps not too surprising, since the coefficients are vacuum-to-excited-state matrix elements of the nucleon interpolating fields. Relevant in practice is that the low-energy-coefficients (LECs) associated with the interpolating fields, which are completely unknown, do not enter at LO in the chiral expansion. Therefore, LO ChPT makes a definite prediction for the multi-particle-state contribution to the 2pt function and the effective mass.   

An alternative way to deal with multi-particle states in spectroscopy calculations is the well-known variational method \cite{Luscher:1990ck}. It can be used provided interpolating fields for the multi-particle states are taken into account.\cite{Lang:2012db,Kiratidis:2015vpa,Lang:2016hnn,Kiratidis:2016hda} However, the numerical cost grows significantly with the number of interpolating fields since the generalized eigenvalue problem one has to solve gets larger. In addition, the number of Wick contractions involved in computing the correlation functions grows rapidly once interpolating fields for multi-particle states are used.\cite{Lang:2016hnn}

So far we discussed the simplest nucleon correlation function, the 2pt function. The impact of excited states is expected to be even more severe in 3pt functions, which are usually measured to calculate various nucleon charges, for example. The nucleon charge $g_X$ refers to the matrix element $\langle N(\vec{p})| O_X |N(\vec{p})\rangle$, where $O_X$ denotes either the vector or axial vector current, the tensor or scalar density. The most prominent one is certainly the nucleon axial charge $g_A$. It is very precisely known experimentally from neutron beta decay, $g_{A,{\rm exp}}=1.2723(23)$ \cite{Olive:2016xmw}. 
The scalar and tensor charge  $g_S, g_T$ are only poorly known, but there is a revived interest in them: 
New nuclear beta-decay experiments aim at an order of magnitude more precise upper bounds for these charges. In order to constrain beyond-standard-model physics lattice QCD estimates for these charges with 10-15\% uncertainties are needed.\cite{Bhattacharya:2011qm}

The standard way to compute the nucleon charges is well-known: One first computes the 3pt function 
\begin{equation}\label{Def3pt}
G_{{\rm 3pt},X}(t,t') = \int d^3x\int d^3y \,\Gamma'_{X,\alpha\beta} \langle N_{\beta}(\vec{x},t) O_{X}(\vec{y},t') \overline{N}_{\alpha}(\vec{0},0)\rangle\,
\end{equation}
involving the same interpolating nucleon fields as in the 2pt function. The operator $O_{X}(\vec{y},t')$ is placed at an operator insertion time $t'$ satisfying $t> t'>0$. $\Gamma'_{X}$ denotes a spin projection matrix chosen such that the leading part of the 3pt function is proportional to the nucleon charge $g_X$ (see Ref.\ \citen{Abdel-Rehim:2015owa}, for instance). Secondly, one forms the ratio with the 2pt function,
\begin{eqnarray}\label{Defratiogeneric}
R_X(t,t')=\frac{G_{{\rm 3pt},X}(t,t')}{G_{\rm 2pt}(t)}\,.
\end{eqnarray}
Performing the standard spectral decomposition of the two correlation functions and taking all times $t,t'$ and $t-t'$ to be large it is straightforward to show that the ratio $R_X$ has the following asymptotic form:
\begin{eqnarray}
\label{DefRatio}
R_X(t,t')= g_X + \sum_n \Big(b_{X,n} e^{-\Delta E_n (t-t')} + \tilde{b}_{X,n} e^{-\Delta E_n t'} + \tilde{c}_{X,n} e^{-\Delta E_n t }\Big).
\end{eqnarray}
By construction, the ratio tends to a constant, the nucleon charge one is interested in. In addition there are excited-state contributions from resonance and multi-particle states. The coefficients $b_{X,n},\tilde{b}_{X,n}$ and $\tilde{c}_{X,n}$ are ratios of various matrix elements involving $O_{X}$ and the interpolating fields $N,\overline{N}$. 

The main point is that in practice we are here confronted with smaller time separations, $t-t'$ and $t'$, than in the 2pt function. Therefore, we expect a less efficient exponential suppression and larger excited-state contributions in the calculation of the nucleon charges. How large the excited-state contribution actually is depends also on the size of the coefficients $b_{X,n},\tilde{b}_{X,n}$ and $\tilde{c}_{X,n}$. Note that not even the sign of the coefficients is a priori determined, and depending on the signs the excited-state contamination can result in $R_X(t,t')> g_X$ or $R_X(t,t')< g_X$.

As for the 2pt function, ChPT can be employed to compute the coefficients associated with the $N\pi$ states, which are expected to contribute the dominant multi-particle-state contribution for large time separations.
To LO the results for these coefficients are essentially parameter free predictions, i.e.\ no unknown LECs enter the results. In particular, the LECs associated with the nucleon interpolating fields do not enter at LO. In the end we are able to obtain concrete estimates for the $N\pi$ contribution in all three charges.  

The calculation is not restricted to the various nucleon charges, but can be repeated with little modification to the calculation of the $N\pi$ contribution to moments of parton distribution functions (PDFs). The reason is that these moments, at leading twist, can be extracted from nucleon matrix elements involving local one-derivative operators. The strategy to compute these in lattice simulations is the same as outlined before for the nucleon charges. The most interesting moments are the the quark momentum fraction $\mf$, the helicity moment $\hm$ and the transversity moment $\tm$. As non-singlet quantities their lattice calculation is fairly straightforward, thus these moments are, together with the nucleon axial, tensor and scalar charge, the simplest hadron structure observables one can measure on the lattice. Recently, the ETM collaboration has presented results for all six observables obtained in physical point simulations \cite{Abdel-Rehim:2015owa}.

\section{ChPT including nucleons}

\subsection{Preliminaries}

ChPT\cite{Weinberg:1978kz,Gasser:1983yg,Gasser:1984gg}, the low-energy effective theory of QCD involving the pseudoscalar fields, can be extended to include heavy matter fields as well.\footnote{There exist numerous introductions to ChPT. Examples are Refs.\ \citen{Colangelo:2000zw,Gasser:2003cg,Golterman:2009kw,Scherer:2012xha,Ecker:2013xja}.}   Probably the most important extension is the inclusion of baryons, leading to so-called baryon ChPT (BChPT) \cite{Gasser:1987rb,Becher:1999he}. The main idea is to consider the baryons as massive matter fields that couple to pions weakly as long as the exchanged 3-momenta are small. This also implies that baryon number is conserved in all interaction processes. 
Moreover, the baryon-pion coupling is constrained by spontaneously broken chiral symmetry. 

Baryon ChPT has a variety of applications, for instance the computation of nucleon-pion scattering or nucleon form factors. Having lattice QCD in mind, BChPT has also been used to compute the quark mass dependence of nucleon observables, and FV effects due to the pions. A review of BChPT that also covers applications to lattice QCD is given in Ref.\ \citen{Bernard:2007zu}. 
In addition there exist many useful lecture notes which differ in scope and length, see Refs.\ \citen{Kubis:2007iy,Scherer:2009bt,Scherer:2012xha}, for instance. The reader is referred to these resources for introductions to BChPT. Here we will focus only on those aspects that are relevant for the application we have in mind, the calculation of the $N\pi$ contamination in nucleon correlation functions measured in lattice QCD. 

\subsection{Chiral Lagrangian and effective operators to LO}\label{sect:BChPT}

ChPT is an expansion in powers of pion momenta and quark masses. Thus, the chiral lagrangian is organized in powers of derivatives acting on the pion fields and in powers of the light quark masses. The standard power counting rules count two derivatives as one quark mass \cite{Gasser:1983yg}. In addition to the power counting the main construction principle for the chiral lagrangian rests on symmetries: One simply writes down the most general chiral Lagrangian that is compatible with the symmetries of QCD, the underlying theory ChPT is supposed to describe at low energies. 

In the following we use the covariant formulation of BChPT for reasons of convenience. Manifest Lorentz covariance simplifies the construction of the effective operators needed for our calculation, as discussed below. In addition, many useful results can be found in the literature and need not be derived from scratch. For simplicity we consider SU(2) BChPT with exact isospin symmetry. Therefore, the degrees of freedom in the chiral theory are the three mass degenerate pions and the mass degenerate proton and nucleon, forming the nucleon doublet.

The chiral Lagrangian to LO has been known for quite some time \cite{Gasser:1987rb} and reads
\begin{equation}\label{effLag}
{\cal L}_{\rm eff}={\cal L}_{N\pi}^{(1)} + {\cal L}_{\pi\pi}^{(2)}\,.
\end{equation}
Here ${\cal L}_{\pi\pi}^{(2)}$ is the standard two-flavor mesonic chiral Lagrangian to LO \cite{Gasser:1983ky}. 
The other part ${\cal L}_{N\pi}^{(1)}$ contains the nucleon fields and their coupling to the pions. The superscripts denote the low-energy (or chiral) dimension of these lagrangians, i.e.\ they count the number of derivatives in these terms \cite{Gasser:1987rb}. In mesonic ChPT the Lagrangians come only in even powers of derivatives, simply because an even number of derivatives is needed to construct a Lorentz scalar. The nucleon-pion part, on the other hand, allows even and odd powers, essentially due to the possible presence of Dirac matrices in terms containing the (fermionic) nucleon field. 

Expanding ${\cal L}_{\rm eff}$  in powers of pion fields and keeping interaction terms with one pion field only we find (for euclidean space time)
\begin{eqnarray}
\label{Leff}
{\cal L}_{\rm eff} &=& \overline{\Psi} \Big(\gamma_{\mu}\partial_{\mu} +\mN \Big)\Psi +\frac{1}{2}\pi^a \Big(- \partial_{\mu}\partial_{\mu} + M_{\pi}^2 \Big)\pi^a + \frac{ig_A}{2f}\overline{\Psi}\gamma_{\mu}\gamma_5\sigma^a \Psi \, \partial_{\mu} \pi^a\,.
\end{eqnarray}
The  nucleon fields $\Psi=(p,n)^T$ and $\overline{\Psi}=(\overline{p},\overline{n})$ 
contain the Dirac fields for the proton $p$ and the neutron $n$.  $M_{\pi}=2Bm$ and $f$ denote the pion mass and the chiral limit value of the pion decay constant. Both $f$ and $B$ (related to the chiral condensate) are LECs associated with ${\cal L}_{\pi\pi}^{(2)}$. $\mN$ and $g_A$ are the chiral limit values of the nucleon mass and the nucleon axial charge. These are two LECs stemming from ${\cal L}_{N\pi}^{(1)} $.

The third term on the right hand side of \pref{Leff} is the leading interaction term. It is proportional to the axial charge and couples two axial vectors to obtain a Lorentz scalar. It describes the well-known one-pion exchange interaction between two nucleons. 
In the following we will present LO results only, and for those the lagrangian in eq.\  \pref{Leff} will be sufficient. Even though we will not encounter any of the difficulties that appear at higher than LO a short comment seems appropriate. 

The chiral Lagrangian for BChPT is known through ${\cal L}_{N\pi}^{(4)}$ \cite{Fettes:2000gb}. The terms at this order are needed for calculations at the one-loop level. However, the loop-expansion in BChPT is non-trivial, because loop diagrams do not automatically obey the naive power-counting rules of the chiral expansion. The reason is the nucleon mass that enters the loop integrals, which is not small and does not vanish in the chiral limit. 
Possible remedies for this problem are well-known in the literature. For example, heavy-baryon ChPT (HBChPT) \cite{Georgi:1990um,Jenkins:1990jv} eliminates the nucleon mass  mass from the nucleon propagator and is essentially an additional expansion in inverse powers of $\mN$. Another method is the so-called infrared regularization, which is a manifestly covariant way to calculate loop diagrams in BChPT.\cite{Becher:1999he} The main ideas of the various approaches are carefully explained in Ref.\ \citen{Bernard:2007zu}. Even though we do not need the details here since we work at LO only, it is worth emphasizing that the techniques to perform calculations beyond that order are well understood.

We are mainly interested in the 3pt functions involving various quark bilinears, so we need the expressions for these operators in the effective theory. The expressions for the vector and axial vector currents, as well as the scalar density are well-established. They are usually derived from the effective Lagrangian in presence of external source fields for the operators and by taking derivatives with respect to these external fields. For example, for the vector and axial vector currents one finds the following LO expressions,
\begin{eqnarray}
V_{\mu}^a & = & \overline{\Psi}\gamma_{\mu}\sigma^a\Psi -\frac{g_A}{f}\epsilon^{abc} \pi^b \overline{\Psi}\gamma_{\mu}\gamma_5 \sigma^c\Psi - 2i \epsilon^{abc}\partial_{\mu}\pi^b\pi^c\,,\label{DefVector}\\
A_{\mu}^a & = & g_A  \overline{\Psi}\gamma_{\mu}\gamma_5 \sigma^a\Psi -\frac{1}{f}\epsilon^{abc} \pi^b \overline{\Psi}\gamma_{\mu}\sigma^c\Psi - 2i f\partial_{\mu} \pi^a\,.\label{DefAxial}
\end{eqnarray}  
The first two terms in each expression on the right hand side stem from ${\cal L}_{N\pi}^{(1)}$, the remaining one from ${\cal L}_{\pi\pi}^{(2)}$. Each term in these expressions transforms correctly as a vector or axial vector current. Still, the LECs associated with each term are not arbitrary but involve only the LECs we have already found in the effective Lagrangian \pref{Leff}. This is a consequence of the chiral Ward identities of QCD. Chiral symmetry relates correlation functions involving the vector and axial vector currents, and the effective theory reproduces the Ward identities only if the LECs in the chiral Lagrangian and the expressions for the operators are related.

Similar relations can also be observed in the expressions for the operators associated with the PDF moments. At the quark level the quark momentum fraction $\mf$ and the helicity moment $\hm$ can be extracted from the nucleon matrix elements of the operators
\begin{equation}
\OV{\mu\nu} \,=\, \overline{q}\gamma_{{\{\mu}}D^-_{\nu\}} T^a q\,,\qquad
\OA{\mu\nu} \,=\, \overline{q}\gamma_{{\{\mu}}D^-_{\nu\}}\gamma_5 T^a q\,.
\end{equation}
$q=(u,d)^T$ denotes the isospin quark doublet, the SU(2) generators are defined as half of the Pauli matrices, $T^a=\sigma^a/2$, 
 and the (color covariant) derivative is given by
$D^-_{\mu} = ( {\overrightarrow D}_{\mu} -  \overleftarrow{D}_{\mu})/2$.
The curly brackets refer to symmetrization with respect to the two Lorentz indices, and a subtraction of the trace part is also implied. 

Based on the transformation properties under chiral symmetry, parity and charge conjugation the ChPT expressions for these  operators have been constructed in Refs.\ \citen{Dorati:2007bk,Wein:2014wma}. To LO and to one power in the pion fields one finds the expressions
\begin{eqnarray}
\OV{\mu\nu} &=& a_{2,0}^v \overline{\Psi} \gamma_{\{\mu}\partial_{\nu\}}^- \sigma^a\Psi -  \frac{\Delta a_{2,0}^v}{f} \epsilon^{abc} \pi^b \overline{\Psi} \gamma_{\{\mu}\gamma_5\partial_{\nu\}}^- \sigma^c\Psi\,,\label{OpVector}\\
\OA{\mu\nu} &=& \Delta a_{2,0}^v  \overline{\Psi} \gamma_{\{\mu}\gamma_5\partial_{\nu\}}^- \sigma^a\Psi -  \frac{a_{2,0}^v}{f} \epsilon^{abc} \pi^b \overline{\Psi} \gamma_{\{\mu}\partial_{\nu\}}^- \sigma^c\Psi\,.\label{OpAxialVector}
\end{eqnarray}
The derivative $\partial_{\mu}^{-} = ({\overrightarrow \partial}_{\mu} - \overleftarrow{\partial}_{\mu})/2$ contains the standard partial derivatives acting on the nucleon fields. Besides  $f$ these expressions contain two more LECs, $a_{2,0}^v$ and $\Delta a_{2,0}^v$.\footnote{We follow the notation introduced in Ref.\ \citen{Dorati:2007bk}.}  Their normalization is chosen such that these coefficients correspond to the chiral limit values of the momentum fraction $\mf$ and the helicity moment $\hm$, respectively. The  results resemble the expressions for the vector and axial vector currents. There are two contributions and their LECs are related due to chiral symmetry. On the other hand, there is no contribution involving pion fields only. The reason is that Lorentz indices in terms with pion fields can only come from partial derivatives, and we need at least two of those to form a symmetric tensor. Such an expression is necessarily two orders higher in the chiral power counting \cite{Dorati:2007bk}. 

The remaining expression we need for the computation of the scalar and tensor charges $g_S,g_T$ are
\begin{equation}\label{Opscaltensorcharge}
S^a \,=\, g_S \overline{\psi} \sigma^a\psi\,,\qquad T^a_{\mu\nu} = g_T \overline{\psi}\sigma_{\mu\nu}\sigma^a\psi\,,
\end{equation}
and the operator used to obtain the transversity moment $\tm$ is given by\footnote{Square brackets refer to antisymmetrization with respect the associated Lorentz indices.} 
\begin{equation}\label{OpTensor}
\OT{\mu\nu\rho} \,=\, \delta a_{2,0}^v \overline{\Psi} \sigma_{[\mu \{\nu ]}\partial_{\rho\}}^- \sigma^a\Psi \,.
\end{equation}
Here too we have already expanded in powers of pion fields, and we dropped all contributions involving more than one. The LEC $\delta a_{2,0}^v$ in \pref{OpTensor}  is chosen such that it corresponds to the chiral limit value of the transversity moment. 

All these expressions consist of one term only at LO. Other contributions are necessarily of higher order, for instance purely pionic contributions to the tensor operators since these involve too many partial derivatives acting on the pion fields. 

In their original papers (Refs.\ \citen{Gasser:1983yg,Gasser:1984gg}) Gasser and Leutwyler did not consider source fields for the tensor bilinear $\overline{q} \sigma_{\mu\nu}T^a q$. 
A reason might be that in  the standard model there is no obvious source that couples directly to it, in contrast to the vector and axial vector currents that couple to the electromagnetic and weak interactions. 
Still, tensor source fields can be included along the same lines as source fields for the vector and axial vector currents or the scalar and pseudo scalar densities. Taking into account the symmetry properties under chiral symmetry, parity and charge conjugation the source term is mapped to ChPT and the effective operator is obtained as usual via a derivative with respect to the source field.    
Mesonic ChPT with a tensor source field has been constructed some time ago in Ref.\ \citen{Cata:2007ns}, but the generalization to covariant BChPT is, to our knowledge, missing. Following the general construction steps mentioned before the LO chiral expressions for the tensor operators in eqs.\ \pref{Opscaltensorcharge} and \pref{OpTensor} are derived in Refs.\ \citen{Bar:2016uoj,Bar:2016jof}. 

Finally, we need the chiral expressions for the nucleon interpolating fields entering the correlation functions. Based on the symmetry properties of the interpolating fields on the quark level we construct the most general expression in the effective theory that has the same symmetry properties. 

At the quark level there exist many choices for the interpolating field $N$ with the quantum numbers of the nucleon. 
The number is significantly reduced if we consider local operators composed of three quark fields at the same point $x$.
If, in addition, we constrain ourselves to operators  without derivatives there exist only five different ones, but, as a consequence of Fierz identities, only two of them are independent \cite{Ioffe:1981kw,Espriu:1983hu,Nagata:2008zzc}. 
To write them down it is convenient to introduce the quark field doublet $\tq$ as 
\begin{equation}\label{qtilde}
\tq=q^{\rm T} C \gamma_5 (i\sigma_2)\,,
\end{equation}
with  $C$ denoting the charge conjugation matrix. With this definition the two independent nucleon operators can be written as
\begin{equation}\label{DefN12}
 N_1 \,= \, (\tq q) q\,,\qquad
 N_2\,=\, (\tq \gamma_5 q) \gamma_5q\,.
\end{equation}
This compact form suppresses the contraction of the isospin and Dirac indices in the bilinear quark fields $(\tq q)$ and $(\tq \gamma_5 q)$ (``diquarks'') and the summation over the color indices with the totally antisymmetric tensor  $\epsilon_{abc}$ to form a color singlet. 
Note that the nucleon operators $N_i$ are still isospin doublets. To project onto the quark content of the proton and neutron one has to contract with the isospin basis vectors $e_p=(1,0)^{\rm T}$ and $e_n=(0,1)^{\rm T}$, respectively. 

The transformation properties of the $N_k, k=1,2$ under chiral and parity transformations are easily worked out from the transformation behavior of the quark field, and the details can be found in Ref. \citen{Nagata:2008zzc}. The mapping to BChPT is then straightforward and has essentially been done in Ref.\ \citen{Wein:2011ix}. At lowest chiral dimension (which is zero) only one effective operator with one unknown LEC $\tilde{\alpha}_k$ contributes. Expanding it in powers of pion fields and keeping only the terms up to linear order we obtain for $N_k$ the chiral expression
\begin{equation}\label{Neffexp}
N_k(x)=  \tilde{\alpha}_k \left(\Psi(x) + \frac{i}{2f} \pi^c(x) \sigma^c\gamma_5\Psi(x)\right)\,,\qquad k=1,2\,.
\end{equation}
The first term is proportional to the nucleon field $\Psi$, as expected. The second term involves a nucleon-pion coupling
that will contribute to the two-particle $N\pi$ contribution in the correlation functions we are interested in. 

Note that the chiral expression \pref{Neffexp} is essentially the same for the two interpolating fields, the only difference being the a priori different values for the LECs $\tilde{\alpha}_1$ and $\tilde{\alpha}_2$. The reason is that the two interpolating fields in eq.\ \pref{DefN12} transform the same way under chiral and parity transformations, thus both map onto the same operator in the effective theory. 

Even though not directly needed in the following we also quote the chiral expressions for the interpolating fields at the next order. 
At chiral dimension 1 two terms with two LECs $\tilde{\beta}_{k,1}$ and $\tilde{\beta}_{k,2}$ contribute.\cite{Wein:2011ix}
 Expanding again to linear order in the pion fields we obtain\footnote{For simplicity we use a slightly different notation for the LECs than Ref.\ \citen{Wein:2011ix}.}
\begin{equation}\label{Neffcd1}
N_k^{(1)} =  \frac{\tilde{\beta}_{k,1}}{f} \partial_{\mu}\pi^c\sigma^c\gamma_{\mu}\gamma_5 \Psi +  \frac{\tilde{\beta}_{k,2}}{f} \partial_{\mu}\pi^c\sigma^c\gamma_5 \partial_{\mu}\Psi\,.
\end{equation}
In contrast to the $N\pi$ term in \pref{Neffexp} the pion fields enter with an additional partial derivative, thus the terms in \pref{Neffcd1} have chiral dimension 1.  

So far we considered local interpolating fields only. In lattice QCD so-called smeared interpolators are very often used to suppress excited-state contributions in the correlation function. Smeared nucleon interpolating fields are formed as in \pref{DefN12} but with the local quark fields replaced by smeared ones. These are generically of the form
\begin{equation}
q_{\rm sm} (x) = \int {\rm d^4}y K(x - y) q(y)
\end{equation}
with some gauge covariant kernel $K(x-y)$ which is essentially zero for $|x-y|$ larger than some ``smearing radius'' $R_{\rm sm}$. The kernel depends on the details of the smearing procedure. For example, Gaussian and exponential smearing \cite{Gusken:1989ad,Gusken:1989qx,Alexandrou:1990dq} is local in time and the kernel contains a delta function in the euclidean time coordinate. In contrast, the gradient flow \cite{Luscher:2013cpa} is a truly four-dimensional smearing. 

What matters for the mapping to ChPT are the transformation properties of the smeared quark fields. Provided the kernel is diagonal in spinor space (as it usually is, for example for the smearing methods mentioned before) the smeared quark fields transform just as the unsmeared ones under parity and global chiral transformations.  Consequently, also the nucleon interpolating fields formed with the smeared quark fields transform just as their local counterparts. Since the symmetry properties of the interpolating fields determine their expression in ChPT we can conclude that both local and smeared interpolating fields are mapped onto the same effective operator. The only difference are the different values for the LECs.

However, one condition has to be imposed here. Smeared interpolators with some ``size'' are mapped onto pointlike nucleon field in the chiral effective theory. For this to be a good approximation the smearing radius needs to small compared to the Compton wave length of the pion\cite{Luscher:2013vga},
\begin{equation}\label{localitybound}
R_{\rm sm} \ll \frac{1}{M_{\pi}}\,.
\end{equation}
Provided this condition is met the pions do not distinguish between smeared and pointlike interpolating fields. A concrete example is given in Ref.\ \citen{Bar:2013ora} where ChPT for observables based on the gradient flow has been constructed.

For physical pion masses the right hand side of \pref{localitybound} is about 1.4 fm. For smearing radii of a few tenths of a fermi the bound seems reasonably well satisfied. In practice $R_{\rm sm}$ of 0.5 fm or even larger are often used, which are probably too large for the description of smeared interpolators by point like nucleon fields. Eventually this has to be checked by comparing the results of ChPT with actual lattice data. 

We emphasize that the presence of $N\pi$ terms in \pref{Neffexp} and \pref{Neffcd1} has nothing to do with smearing, these contributions are also present for point like interpolating fields. Smearing effects the values of the LECs in these expressions. 

\section{The correlation functions in ChPT}

\subsection{Preliminaries}

The 2pt and 3pt functions in eqs.\ \pref{Def2ptfunc}, \pref{ExpAnsatz} can be computed perturbatively in ChPT provided the time scales $t$ and $t'$ are large. 
Large here means that the correlation functions are dominated by pion physics that is captured by ChPT. 

Given the explicit expressions in the last section, eqs.\ \pref{Leff} - \pref{OpTensor} and eq.\ \pref{Neffexp}, the perturbative expansion of the correlation functions is standard and straightforward. Recall that we assume a finite spatial volume with periodic boundary conditions imposed, so the spatial momenta in the propagators are discrete. For simplicity we assume an infinite time extent leading to a simple exponential decay of the correlation functions. 

Since we are interested in the time dependence of the correlation functions the time-momentum representation of the propagators is convenient. The pion propagator, for instance, reads 
\begin{equation}\label{scalprop}
G^{ab}_{\pi}(x,y)=   \delta^{ab}L^{-3}\sum_{\vec{p}_n} \frac{1}{2 E_{\pi,n}} e^{i\vec{p}_n(\vec{x}-\vec{y})} e^{-E_{\pi,n} |x_0 - y_0|}
\end{equation} 
in this representation, with pion energy $E_{\pi,n} =\sqrt{\vec{p}_n^2 +\mpi^2}$. The sum runs over the spatial momenta $\vec{p}_n$ allowed by the boundary conditions. An analogous expression is found for the nucleon propagator \cite{Bar:2015zwa}. 

\begin{figure}[tb]
\begin{center}
\includegraphics[scale=0.45]{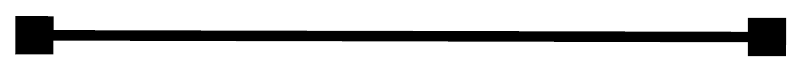}\hspace{1cm}\includegraphics[scale=0.45]{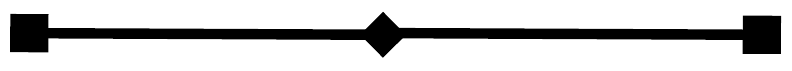}\\
a)\hspace{4.4cm} b)
\caption{Leading Feynman diagrams for the 2pt and 3pt functions. The squares represent the nucleon interpolating field at times $t$ and $0$, the diamond stands for the operator insertion at time $t'$. Solid lines represent the nucleon propagator.} 
\label{fig:Nfeynmandiagrams}
\end{center}
\end{figure}

The leading Feynman diagrams are depicted in figure \ref{fig:Nfeynmandiagrams}. They are essentially the (zero-momentum) nucleon propagator and provide the leading single-nucleon-state contribution to the correlation functions. For example, the result for the 2pt function is given by\cite{Bar:2015zwa}
\begin{equation}\label{SingleNucl2pt}
G^{N}_{{\rm 2pt}}(t) =2|\tilde{\alpha}_k|^2 e^{-M_Nt}\,. 
\end{equation}   
Comparing this result with the general expression in eq.\ \pref{ExpAnsatz} we indeed find the first exponential and identify $c_0$ with  $2|\tilde{\alpha}_k|^2$.
Note that for \pref{SingleNucl2pt} we have taken the same interpolating field at source and sink. This implies that the LECs associated with the interpolating fields appear as $|\tilde{\alpha}_k|^2$ and $G^{N}_{{\rm 2pt}}$ is real and positive. If different interpolating fields are used at source and sink this will not be the case: The LECs appear as the product $\tilde{\alpha}_k\tilde{\alpha}_{k'}^*$ and the reality and positivity properties of $G^{N}_{{\rm 2pt}}$ are lost. 

The single-nucleon contributions to the 3pt function are also very simple: The 3pt function is proportional to the 2pt function, and the proportionality constant is essentially either the charge or the moment associated with the operator used in the 3pt function. For example, in case of the axial vector current we find 
\begin{equation}\label{SingleNucl3ptA}
G^{N}_{{\rm 3pt},A}(t,t') = g_A G^{N}_{{\rm 2pt}}(t). 
\end{equation} 
Thus, we confirm the asymptotic value for the ratio $R_A$ as specified in eq.\ \pref{DefRatio}. 
Note that this result and the analogous ones for the other operators are obtained by construction. They essentially mean that the identification of the unknown LECs in eqs.\  \pref{DefVector} - \pref{OpTensor} with the charges and moments has been done correctly.

\subsection{$N\pi$ contribution in the 2pt function}

\begin{figure}[tb]
\begin{center}
\includegraphics[scale=0.45]{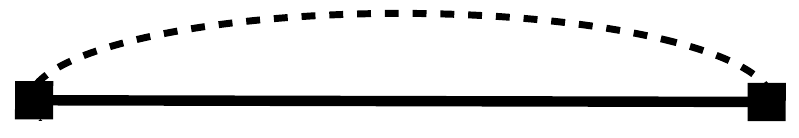}\hspace{1cm}\includegraphics[scale=0.45]{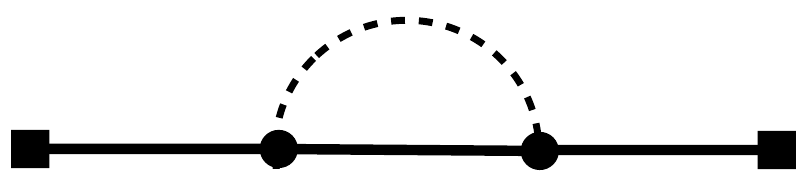}\\
a)\hspace{4.4cm} b)\\[3ex]
\includegraphics[scale=0.45]{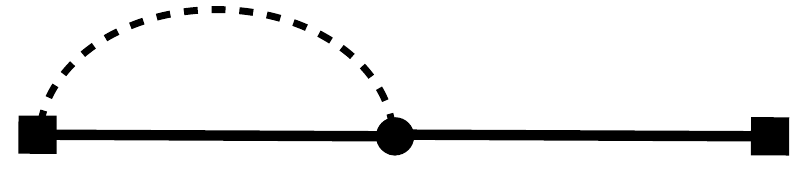}\hspace{1cm}\includegraphics[scale=0.45]{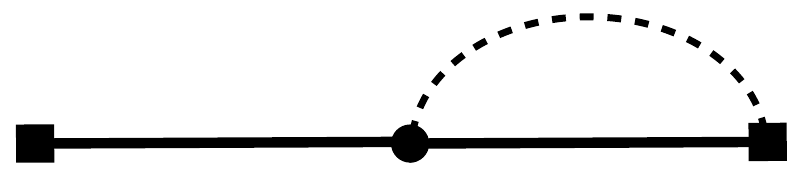}\\
c)\hspace{4.4cm} d)
\caption{Feynman diagrams for the nucleon correlation function. The squares represent the nucleon interpolating fields at times $t$ and $0$. The circles represent a vertex insertion at an intermediate space-time point; and an integration over this point is implicitly assumed. The solid and dashed lines represent nucleon and pion propagators, respectively. }
\label{fig:Npidiagrams2pt}
\end{center}
\end{figure}

Figure \ref{fig:Npidiagrams2pt} shows the leading diagrams with an $N\pi$-state contribution to the 2pt function, i.e.\ they contain a contribution that drops off exponentially with the total energy $E_{N\pi,n} = E_{N,n}+E_{\pi,n}$ of a nucleon-pion state, where the spatial momenta are back-to-back, $\vec{p}_{N,n}=-\vec{p}_{\pi,n}$. We are interested in the prefactor of the exponential $\exp(-E_{N\pi,n}t)$, which is easily read off once the diagram is computed. 

We emphasize that even though the diagrams in fig.\ \ref{fig:Npidiagrams2pt} are 1-loop diagrams the calculation of the $N\pi$-state contribution is a tree-level calculation. However, diagrams b) - d) also contain single-nucleon contributions dropping off with $\exp(-M_N t)$. We can ignore these contributions for our purposes here, at least in a LO calculation, but for this single-nucleon contribution the diagrams do involve a sum over all momenta of the intermediate $N\pi$ pair. Diagram b) is a familiar example. This diagram is essentially the nucleon self-energy diagram that leads to the well-known $M_{\pi}^3$ term in the 1-loop ChPT result for the nucleon mass \cite{Gasser:1987rb}.

The final result for the coefficients in the 2pt function illustrates some generic features, and it is simple enough to be displayed completely: 
\begin{eqnarray}
C^{N\pi}_{{\rm 2pt}}& =&2|\tilde{\alpha}_k|^2 \sum_{\vec{p}_n} c_{{\rm 2pt},n} e^{-E_{N\pi,n} t}\,,\label{NpiNucl2pt}\\
 c_{{\rm 2pt},n} &=& \frac{1}{16(fL)^2E_{\pi,n}L} \left(1-\frac{M_N}{E_{N,n}}\right)C_{{\rm 2pt},n}.\label{c2pt}
\end{eqnarray}
We find it useful to write the coefficients $c_{{\rm 2pt},n}$ as a product of two universal factors and a``reduced" coefficient $C_{{\rm 2pt},n}$. 
The first factor on the right hand side shows the anticipated volume suppression factor $1/L^3$ of a two-particle state in a finite spatial volume. It combines with the pion decay constant and the pion energy to the dimensionless combination $1/(fL)^2E_{\pi,n}L$. 

We already mentioned that the nucleon-pion state with both particles at rest does not contribute to the correlation function because it is parity odd. The second factor $(1-M_N/E_{N,n})$ guarantees exactly this since it vanishes if the momentum of the nucleon (and the pion) is zero. 

The non-trivial result of the ChPT calculation is the remaining term $C_{{\rm 2pt},n}$, which is given by\cite{Bar:2015zwa}
\begin{equation}\label{redcoeff2pt}
C_{{\rm 2pt},n} \,=\, 3\left(\gAbn -1\right)^2\,.
\end{equation}
Here we introduced the (momentum-dependent) combination 
\begin{equation}
\gAbn=g_A\frac{E_{N\pi,n} + M_N}{E_{N\pi,n} -M_N}\,.
\end{equation}
The factor 3 in eq.\ \pref{redcoeff2pt} stems from $N_f^2-1$ with $N_f=2$ in the two-flavor theory.

Comparing these results with the coefficients in the general expression \pref{ExpAnsatz} we find them to be given by $2|\tilde{\alpha}_k|^2c_{{\rm 2pt},n}, \,n>0$. In particular, the LECs associated with the nucleon interpolating fields enter as a multiplicative constant. This can be traced back to eq.\ \pref{Neffexp} where the LEC $\tilde{\alpha}_k$ appears as an overall constant too.
This implies that the LEC combination $2|\tilde{\alpha}_k|^2$ cancels in the ratios $c_n/c_0$ that appear in the result for the effective mass, eq.\ \pref{Meff}. Explicitly we find ($\Delta E_n = E_{N\pi,n} -M_N$)
\begin{equation}
M_{N,{\rm eff}} =  M_N \Bigg( 1+ \sum_{\vec{p}_n} c_{{\rm 2pt},n} \frac{\Delta E_n}{M_N} e^{-\Delta E_n t}\Bigg)\,.\label{Meffexpl}
\end{equation}
Consequently, the LO $N\pi$-state contribution to the effective mass is the same for pointlike and for smeared interpolating fields, since the difference between these interpolators is encoded in their different values for the LECs. 
However, this universality property will be lost at higher orders in the chiral expansion. At the next order diagrams involving the vertices in eq.\ \pref{Neffcd1} need to be taken into account. These contain the LECs $\tilde{\beta}_{k,1}$ and $\tilde{\beta}_{k,2}$, thus the results for the effective mass will depend on the ratios $|\tilde{\beta}_{k,1}/\tilde{\alpha}_k|^2$ and $|\tilde{\beta}_{k,2}/\tilde{\alpha}_k|^2$. 

We already mentioned that the 2pt function was independently calculated in Ref.\ \citen{Tiburzi:2015tta} using HBChPT. In that non-relativistic formulation the antinucleon degrees of freedom are dropped and the dispersion relation of the heavy nucleon is non-relativistic. The HBChPT results can be obtained from the ones derived in the covariant formulation once the appropriate expansion is done: If we expand $E_{N,n}\approx M_N+ p_n^2/2M_N$ in \pref{c2pt} and drop all but the dominant terms we indeed reproduce the result in Ref.\ \citen{Tiburzi:2015tta}.  

We also mention that the LO ChPT results given above were recently rederived in Ref.\ \citen{Hansen:2016qoz} using the Lellouch-L{\"u}scher formalism developed in Ref.\ \citen{Lellouch:2000pv}. Even though the calculational details are quite different compared to the direct computation of the Feynman diagrams in fig.\ \ref{fig:Npidiagrams2pt}, the final result for the 2pt function is the same.

\subsection{$N\pi$ contribution in the 3pt functions}

\begin{figure}[t]
\begin{center}
\includegraphics[scale=0.45]{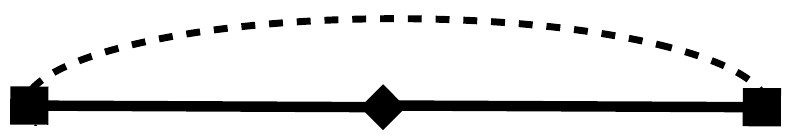}\hspace{0.3cm}\includegraphics[scale=0.45]{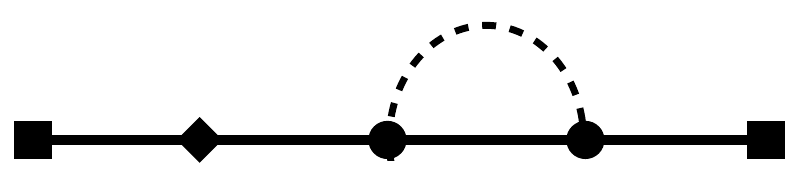}\hspace{0.3cm}\includegraphics[scale=0.45]{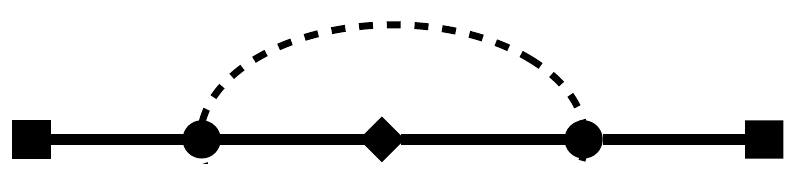}\\ 
a)\hspace{3.5cm} b)\hspace{3.5cm} c)\\[3ex]
\includegraphics[scale=0.45]{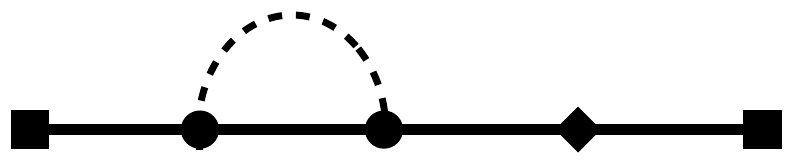}\hspace{0.3cm} \includegraphics[scale=0.45]{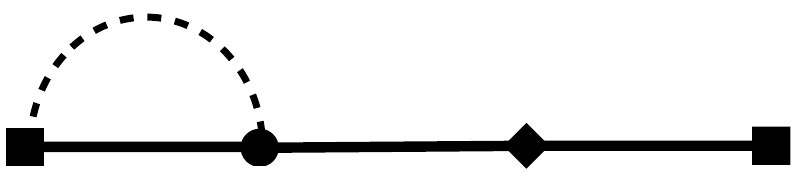}\hspace{0.3cm}\includegraphics[scale=0.45]{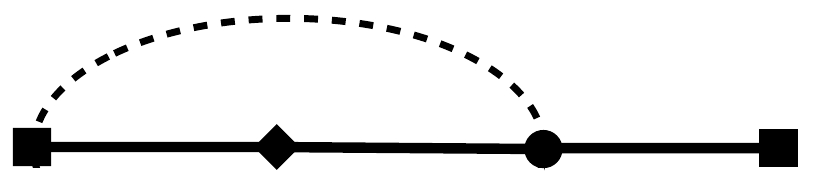}\\
d) \hspace{3.5cm} e)\hspace{3.5cm} f)\\[3ex]
\includegraphics[scale=0.45]{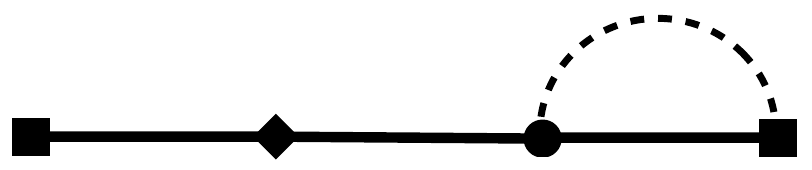}\hspace{0.3cm}\includegraphics[scale=0.45]{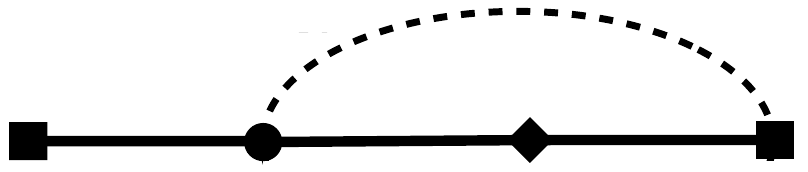}
\includegraphics[scale=0.45]{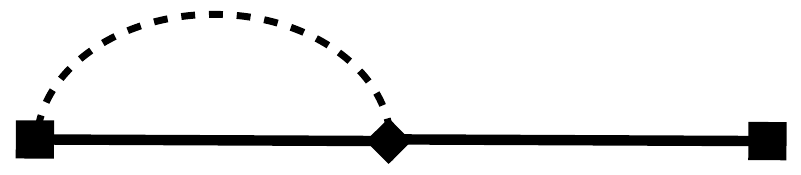}\\ 
g)\hspace{3.5cm} h) \hspace{3.5cm}i)\\[3ex]
\includegraphics[scale=0.45]{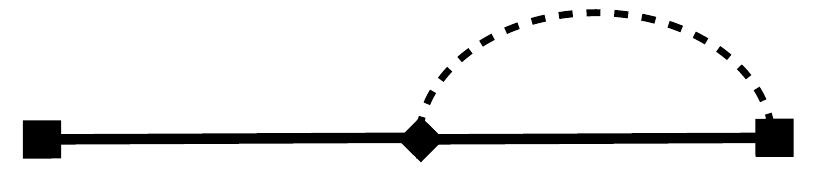}\hspace{0.3cm}\includegraphics[scale=0.45]{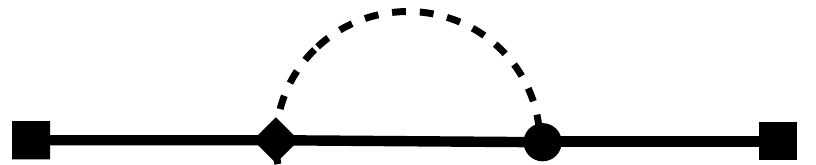}\hspace{0.3cm}\includegraphics[scale=0.45]{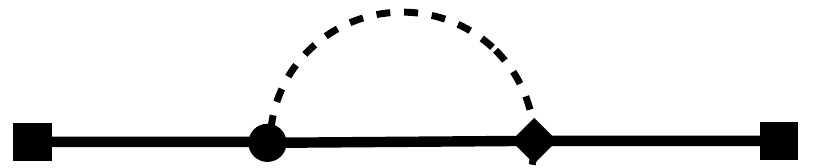}\\
j)\hspace{3.5cm} k)\hspace{3.5cm} l)\\[3ex]
\includegraphics[scale=0.45]{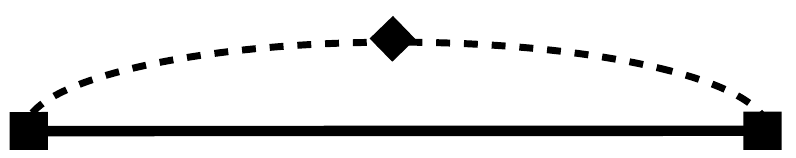}\hspace{0.3cm}\includegraphics[scale=0.45]{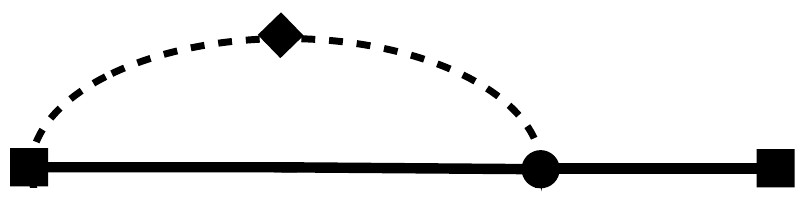}\hspace{0.3cm}\includegraphics[scale=0.45]{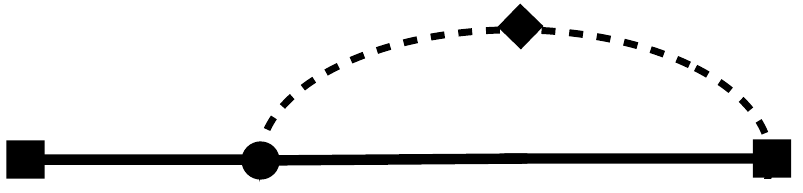}\\
m) \hspace{3.5cm} n)\hspace{3.5cm} o)\\[3ex]
\hspace{0.3cm}\includegraphics[scale=0.45]{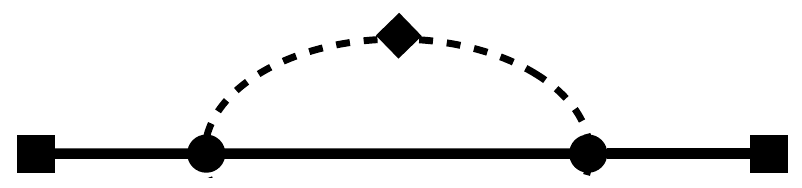}\\
p)
\caption{Feynman diagrams for the LO nucleon-pion contribution in the 3pt functions. Circles represent a vertex insertion at an intermediate space time point, and an integration over this point is implicitly assumed. The dashed lines represent pion propagators.  }
\label{fig:Npidiagrams3pt}
\end{center}
\end{figure}

Figure \ref{fig:Npidiagrams3pt} shows the leading diagrams with an $N\pi$-state contribution to the 3pt functions. The calculation follows the same principles as the corresponding one for the 2pt function. Each diagram has at least one contribution proportional to an exponential $\exp(-E_{N\pi,n}\tau)$, with $\tau$ equal to either $t-t', t'$ or $t$, and we are interested in the prefactors of these  exponentials. As for the 2pt function this is a tree-level calculation despite the loop character of the diagrams in fig.\ \ref{fig:Npidiagrams3pt}.

Not all diagrams contribute for all operators introduced in section \ref{sect:BChPT}. For example, the last four diagrams in fig.\ \ref{fig:Npidiagrams3pt} contribute to the vector current correlator only. They involve the vertex stemming from the pure pion term in eq.\ \pref{DefVector}. For the correlation functions involving the scalar density and the two tensor operators only the first eight diagrams contribute.

Forming the ratio $R_X$ in \pref{Defratiogeneric} the result can be written as
\begin{eqnarray}
\label{DefRatio2}
R_X(t,t')= g_X \Big[1+ \sum_{\vec{p}_n} \Big(b_{X,n} e^{-\Delta E_n (t-t')} + \tilde{b}_{X,n} e^{-\Delta E_n t'} + \tilde{c}_{X,n} e^{-\Delta E_n t }\Big)\Big].
\end{eqnarray}
This is the form of eq.\ \pref{DefRatio}, although the coefficients $b_{X,n}, \tilde{b}_{X,n}, \tilde{c}_{X,n}$ differ by a factor $g_X$.  
The coefficients for the ratios involving the operators we are interested in were computed in Refs.\ \citen{Bar:2016uoj,Bar:2016jof}. The full results are not very illuminating and not shown here. We restrict ourselves to a few comments. 

As in the 2pt function the results can be written as products of a reduced coefficients and the two universal factors also present in the result  for the 2pt function in eq.\ \pref{c2pt}. The factor $(1-M_N/E_{N,n})$ guarantees again that the $N\pi$ state with both the nucleon and pion at rest does not contribute.

By definition the coefficients $b_{X,n}, \tilde{b}_{X,n}, \tilde{c}_{X,n}$ in \pref{DefRatio2} are dimensionless. Thus, they are functions of dimensionless combinations of the various input parameters. In total there are five different ones and a possible choice is
\begin{equation}\label{inputparameters}
f/M_N,\quad g_A,\quad a_{2,0}^v/\Delta a_{2,0}^v,\quad M_{\pi}/M_N,\quad M_{\pi}L\,.
\end{equation}
Recall that the LECs here are the chiral limit values of the pion decay constant, the axial charge and the ratio $\mf/\hm$. On the other hand, the coefficients do not depend on the LECs associated with the nucleon interpolating fields, since these drop out in the ratio. Therefore, the LO results for the ratios is the same for point-like and smeared interpolators. 

In practice the various nucleon charges and moments are obtained from fits of the expected form in eq.\ \pref{DefRatio2} to numerical lattice data. Without any excited-state contribution one performs fits to constants, and each charge and each moment will be an independent fit parameter. No additional fit parameters enter if the $N\pi$ contribution is included in a simultaneous fit of all ratios and the effective mass.\footnote{Provided the pion mass and decay constant have been determined separately from the pseudo scalar 2pt function.} In that sense LO ChPT makes definite predictions: The LO $N\pi$ contributions are given in terms of the LO single nucleon contributions. 

In case of isospin symmetry the vector current is conserved. This implies that the 3pt function of the vector current is exactly given by the conserved charge times the 2pt function. In terms of the coefficients in eq.\ \pref{DefRatio2} this statement reads $b_{V,n}=\tilde{b}_{V,n}= \tilde{c}_{V,n}=0$. This is indeed what one finds in an explicit calculation. Even though not very interesting physically this result has provided a non-trivial check on the programs written to compute the diagrams for general operators $O_X$. In that context it is worth pointing out that the result for $R_A$ has been confirmed in Ref.\ \citen{Hansen:2016qoz}. In addition, in the HB limit we recover the result in Ref.\ \citen{Tiburzi:2015tta} for the axial vector 3pt function.

\section{Impact on lattice calculations}

\subsection{Preliminaries}

The LO $N\pi$ contribution to the nucleon correlation functions we are interested in depends on a five LECs only, see eq.\ \pref{inputparameters}. Fortunately, exactly these LECs are known rather precisely either from experiment or phenomenologically. Assuming these values in our results we get estimates for the $N\pi$ contribution and can make predictions about the expected impact in lattice simulations. 

The five LECs are the chiral limit values of the pion decay constant, the axial charge and the ratio of the average momentum fraction and the helicity moment.  To LO we can use the experimental values for these LECs, i.e.\ we can set  $\gA=1.27$ and $f=f_{\pi}= 93$ MeV.\cite{Olive:2016xmw} The ratio $a_{2,0}^v/\Delta a_{2,0}^v$ is approximately given by $\mf/\hm=0.165/0.19$ \cite{Alekhin:2012ig,Blumlein:2010rn}. We can ignore the errors in these phenomenological values since they are too small to be significant for our LO estimates.

We are mainly interested in the $N\pi$ contributions at the physical point, so we fix the pion and nucleon mass to their physical values. Here it is sufficient to use the simple values $M_{\pi}=140$ MeV and $M_N=940$ MeV. With these values we fixed all parameters in \pref{inputparameters} except the spatial lattice size $L$. 

The sums in eqs.\ \pref{Meffexpl}, \pref{DefRatio2} run over the discrete spatial momenta that are compatible with the periodic boundary conditions we imposed. Not only for practical reasons we need to restrict the sum to momenta smaller than a maximal value satisfying $p_n \le p_{n_{\rm max}}$. 
Finite volume ChPT is an expansion in $p_n/\Lambda_{\chi}$, where the chiral scale $\Lambda_{\chi}$ is typically identified with $4\pi f_{\pi}$ \cite{Colangelo:2003hf}. The smaller $p_{n_{\rm max}}/\Lambda_{\chi}$ the better the ChPT result for the $N\pi$ contribution taken into account in the sum. In order to be a good approximation for the complete $N\pi$ contribution the truncated part needs to be small and negligible, and this can only be achieved by considering the correlation functions for sufficiently large time separations. We will discuss this in more detail in the next two subsections. 

\begin{table}[t]
\caption{ $E_{N\pi,{n_{\rm max}}}$ and $n_{\rm max}$ and as a function of $p_{n_{\rm max}}/\Lambda_{\chi}$; see main text.}
\begin{center}
\begin{tabular}{l|c|c|c|c|}
\multirow{2}{*}{$\frac{p_{n_{\rm max}}}{\Lambda_{\chi}}$}&  \multirow{2}{*}{$\frac{E_{N\pi,{n_{\rm max}}}}{\rm GeV}$}& \multicolumn{3}{c|}{$n_{\rm max}$ }  \\ 
& & $M_{\pi}L=4$ & $M_{\pi}L=5$ & $M_{\pi}L=6$  \\  \hline
0.3 & $\approx 1.35 $ & 2 & 4 & 5 \\
0.45& $\approx 1.6\phantom{3}$& 5 & 8&12 \\
0.6 & $\approx 1.9\phantom{3}$& 10 & 15&22 
\end{tabular}
\end{center}
\label{table:nmax}
\end{table}

In the following we will compare the $N\pi$ contribution for the three different values for $p_{n_{\rm max}}/\Lambda_{\chi}$ specified in table \ref{table:nmax}. Obviously there is some arbitrariness in choosing these values, but for the smallest one with $p_{n_{\rm max}}/\Lambda_{\chi}=0.3$  one expects a reasonably well behaved chiral expansion. This will not be the case for $p_{n_{\rm max}}/\Lambda_{\chi}=0.6$, the largest value we consider. Still, it turns out that for source-sink separations between 1 and 2 fm one has to include $N\pi$ states with that high momenta to saturate the sums in \pref{Meffexpl}, \pref{DefRatio2}.

Table \ref{table:nmax} also includes the energy $E_{N\pi,{n_{\rm max}}}$ of the $N\pi$ state with back-to-back momenta $\vec{p}_{n_{\rm max}}$. For our smallest momentum bound the energy $E_{N\pi,{n_{\rm max}}}$ seems sufficiently well separated from the energy of the first resonance state, which is expected to be somewhere near 1.45 GeV. In that case we may ignore mixing effects with the resonance state that is not captured in LO ChPT.
The energies for the other two bounds are above the energy of the first resonance. Including $N\pi$ states with such high energies without taking into account the effect of the resonance is an approximation, and the results derived from it need to be interpreted with some care.

Placing a bound $p_{n_{\rm max}}$ on the spatial momenta translates into $n_{\rm max}$, the number of $N\pi$ states taken into account in the sums. This number depends on the spatial lattice size $L$. For comparison we consider three different lattice sizes, $M_{\pi}L=4,5$ and $6$. The smallest value is a common value that many collaborations try to reach or exceed in order to keep FV effects small. The largest value is motivated by the simulation setup of the PACS-CS collaboration \cite{Ishikawa:2015rho}. $n_{\rm max}$ for these three volumes are also given in table \ref{table:nmax}. Apparently, the number of states increases rapidly with larger lattices sizes $L$.

\subsection{Impact on the determination of the nucleon mass}

\begin{figure}[t]
\begin{center}
$M_{\rm eff}(t)/M_N$\\
\includegraphics[scale=0.8]{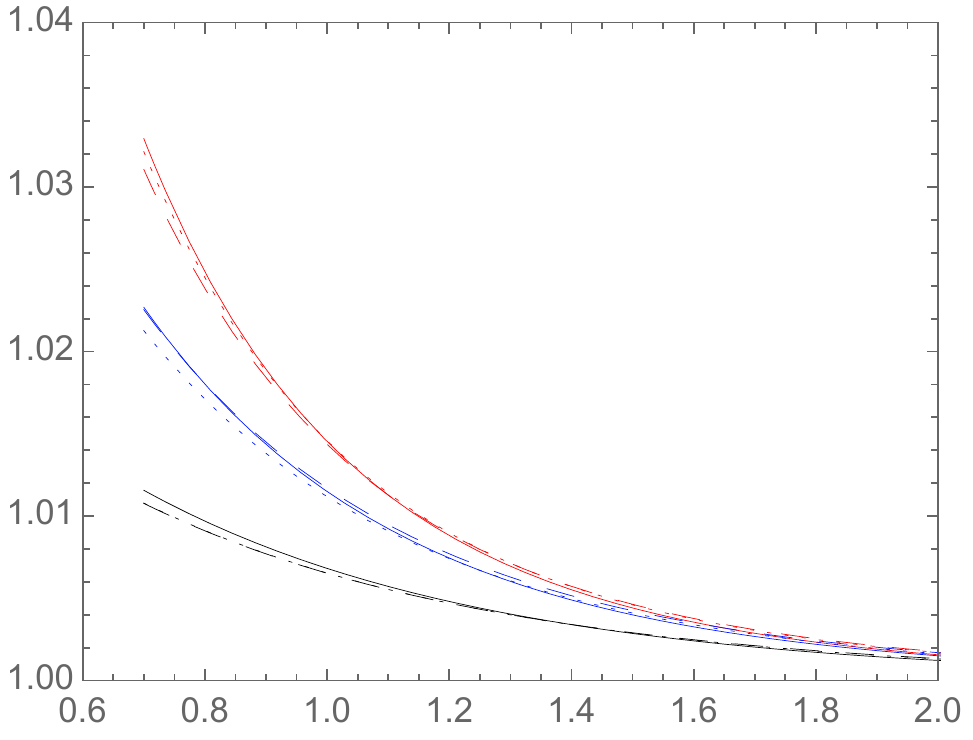}\\
$t$
\caption{The ratio $M_{N,{\rm eff}}(t)/M_N$ as a function of the source sink separation $t$. Results are shown for $M_{\pi}L=4$ (solid lines),  $M_{\pi}L=5$ (dotted lines) and $M_{\pi}L=6$ (dashed lines). Different colors distinguish between the three different energy bounds $E_{N\pi,{n_{\rm max}}}$ specified in table \ref{table:nmax}: first row in black, second row in blue and third row in red.}
\label{fig:effmass}
\end{center}
\end{figure}

Fig.\ \ref{fig:effmass} shows the ratio $M_{N,{\rm eff}}(t)/M_N$ as a function of the source-sink separation $t$. 
Without any excited state contribution this ratio would be equal to 1. Any deviation from this constant value shows directly the $N\pi$ contribution to the effective mass in percent. Results are shown for the three different lattice sizes (solid, dashed and dotted lines)  and the three different energies $E_{N\pi,{n_{\rm max}}}$ (black, blue and red lines) specified in table \ref{table:nmax}. 

\begin{figure}[t]
\begin{center}
$M_{\rm eff}(t)/M_N$\\
\includegraphics[scale=0.8]{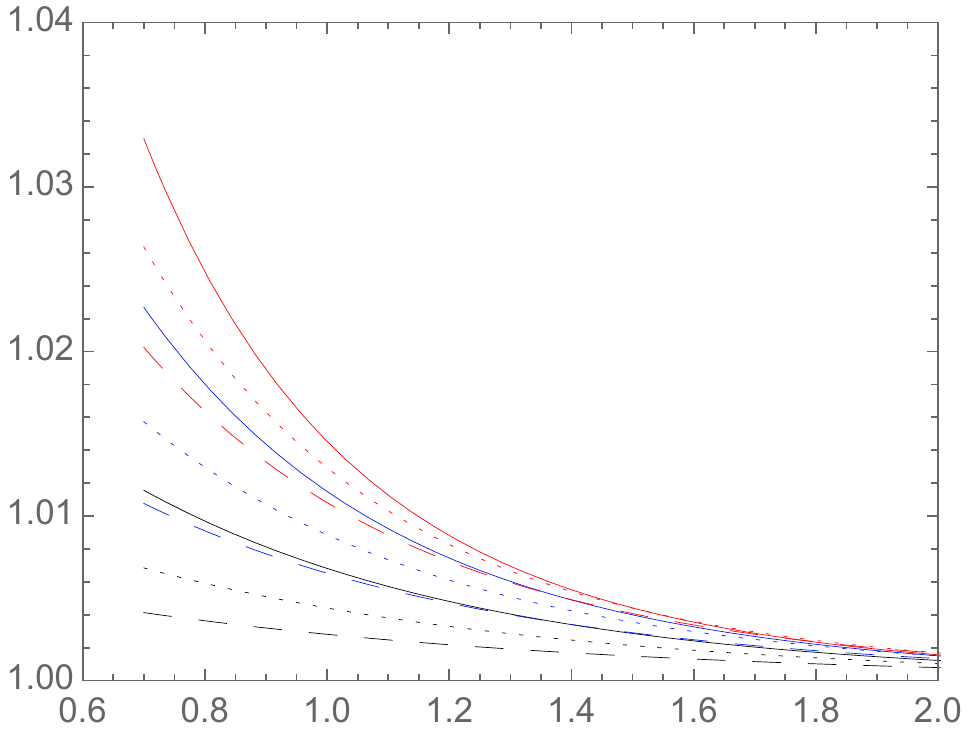}\\
$t$
\caption{The ratio $M_{N,{\rm eff}}(t)/M_N$ as a function of the source sink separation $t$. Results are shown for $M_{\pi}L=4$ (solid lines),  $M_{\pi}L=5$ (dotted lines) and $M_{\pi}L=6$ (dashed lines). Different colors distinguish between three different $n_{\rm max}$ values: 2  (black),  5 (blue) and 10 (red).}
\label{fig:effmassNmaxfixed}
\end{center}
\end{figure}

Apparently, the differences between lines of the same color are very small, i.e.\ the $N\pi$ contribution is essentially the same for the three different $M_{\pi}L$ values displayed in the figure. This stems from the fact that the energy interval of the $N\pi$ states we consider, $[M_N+M_{\pi},E_{N\pi,{n_{\rm max}}}]$, is essentially kept constant for the three different volumes, leading to the different $n_{\rm max}$ values listed in table \ref{table:nmax}. If instead $n_{\rm max}$ is kept fixed as the lattice size $L$ is varied one finds a strong volume dependence. This is illustrated in figure \ref{fig:effmassNmaxfixed}, which shows the results with $n_{\rm max}$ kept fixed at the values for $M_{\pi}L=4$, i.e.\ $n_{\rm max} = 2,5,10$. In that case we can observe a clear spread in the curves with the same color, showing a significant decrease of the $N\pi$ contribution for $M_{\pi}L$ getting larger.

Figure \ref{fig:effmass} also shows the impact of the $N\pi$ states as $n_{\rm max}$ is increased. To a good approximation the red curves saturate the sum in the effective mass; adding more states does not change the results significantly, at least for the source-sink separations larger than about 1 fm. Therefore, we may call the red curve the full $N\pi$ contribution for short.

The figure shows clearly what we remarked before: The larger $t$ the smaller the impact of the high momentum $N\pi$ states relative to the impact of the lowest ones. To be specific let us consider $M_{\pi}L=4$. At $t=1.8$ fm the contribution from the first two states (black curve) makes approximately 75\% of the full contribution (red curve).  At source-sink separations as large as this we may ignore all but the lowest two states. For those we expect our LO result to give a reasonably good estimate for the $N\pi$ contribution; the NLO correction is O($p^2$) and one may expect, as a naive estimate, a 30\% correction. A more honest error estimate requires the result of the NLO calculation.

\begin{figure}[t]
\begin{center}
$M_{\rm eff}(t)/M_N$\\
\includegraphics[scale=0.8]{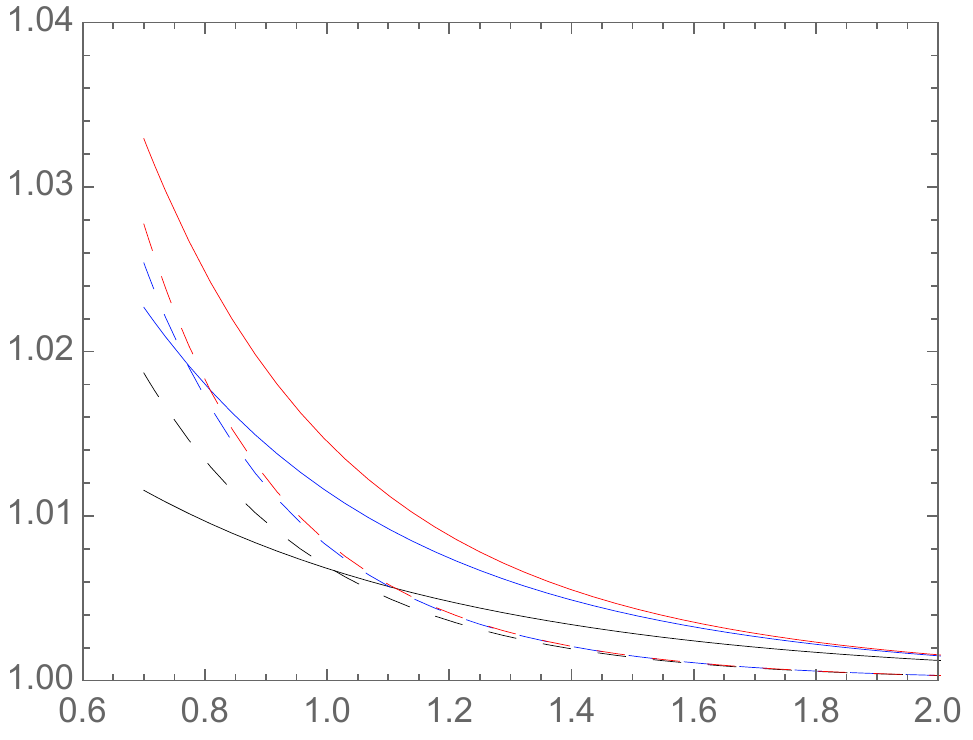}\\
$t$
\caption{The ratio $M_{N,{\rm eff}}(t)/M_N$ as a function of the source sink separation $t$. Results are shown for physical pion masses with $M_{\pi}/M_N =0.145$ (solid lines) and $M_{\pi}/M_N =0.27$ (dashed lines). $M_{\pi}L=4$ in both cases and $n_{\rm max}$ equals  2  (black),  5 (blue) and 10 (red).}
\label{fig:effmassheavypion}
\end{center}
\end{figure}

For smaller $t$ the impact of the higher momentum $N\pi$ states increases slowly. At $t=1.1$ fm the lowest two states contribute approximately 50\% to the full contribution. 
The contribution of the high-momentum $N\pi$ states is prone to larger NLO corrections and we get a cruder estimate. Reading off a +1\%  $N\pi$ contribution at $t\approx1.1$ fm and allowing for a 50\% error due to higher order corrections we arrive roughly at a 1-2\% overestimation of the nucleon mass. The error estimate is a naive guess that can be put on firmer grounds with a calculation at NLO.

For $t$ even smaller than about 1.1 fm the higher momentum $N\pi$ states rapidly dominate the $N\pi$ contribution and we cannot expect the LO ChPT result to be a reasonable approximation anymore. It is also likely that working to higher order in the chiral expansion will not help in going to such small source-sink separations. However, we may still conclude that as many as 10 $N\pi$ states (in case of $M_{\pi}L=4$) contribute substantially to the effective mass for source-sink separations of about 1 fm and below, a slightly unsettling high number.

In any case, the $N\pi$ contribution to the effective mass is rather small for the source-sink separations that are accessible in today's lattice simulations. For $t\gtrsim1.1$ fm it is less than about 1-2\% even if we allow for a 50\% error due to higher order corrections. Whether this systematic uncertainty plays a role in practice depends on the statistical errors in the lattice data. It seems fair to say that the $N\pi$ contribution can be ignored unless the statistical errors in the data are less than a percent. As we will see in the next section, a similarly comforting statement cannot be made for the determination of the various nucleon charges and moments.

Although we are mainly interested in the $N\pi$ contribution for physical pion masses it is interesting to compare the results to those obtained for heavier pions. For a larger than physical pion mass one expects the $N\pi$ contribution to become rapidly smaller. As an illustration, fig.\ \ref{fig:effmassheavypion} shows the ratio $M_{N,{\rm eff}}(t)/M_N$ for $M_{\pi}L=4$
at the physical point (solid lines) compared to those 
for $M_{\pi}/M_N =0.27$ (dashed lines). This value is close to the one found by the 
RQCD collaboration in their simulations with a pion mass of about 295 MeV.\cite{Bali:2014nma} If we keep $M_{\pi}L=4$ fixed the spatial volume is smaller than in the case with a physical pion mass, implying larger discrete spatial momenta of the moving nucleon and pion. Therefore, the energy gaps $\Delta E_n$ in \pref{Meffexpl} are larger and a faster exponential suppression of the $N\pi$ contribution is expected.

Figure \ref{fig:effmassheavypion} supports this expectation. The full $N\pi$ contribution is significantly reduced and the impact of the higher momentum $N\pi$ states is drastically reduced. Even at $t=1.1$ fm the first two states (dashed black curve) contribute approximately 90\% to the full contribution (dashed red curve). For smaller $t$ the contribution of the higher momentum states increases rapidly again. 

Interestingly, the curves for the contribution of the lowest two states (black curves) cross at $t\approx 1$ fm. So despite the larger energy gaps for the heavier pion mass the lowest two states have the same impact as the lowest two states for the physical pion mass. The reason is the larger values for the coefficients $c_{{\rm 2pt},n}$ for the heavier pion mass, which are about a factor   
4 larger than their analogues for the physical pion mass. Even though we need to be careful with drawing conclusions from our LO results at small $t$ values, this example serves as a warning that prejudices about excited-state contributions based on the energy gaps alone can be misleading.

\subsection{Impact on the determination of the charges and moments}

In the following we estimate the impact of the $N\pi$-state contribution on the determination of the various charges and moments. Two methods are widely used in lattice calculations, the plateau and the summation method. Starting point of both methods is the ratio $R_X(t,t')$ defined in eq. \pref{Defratiogeneric}. 

For a given source-sink separation $t$ the $N\pi$-state contribution to $R_X$ is minimal if the operator insertion time $t'$ is in the middle between source and sink. Therefore, the best estimate for the charges and moments is the midpoint value $R_X(t,t/2)$. 
This {\em midpoint estimate} is essentially equivalent to what is called the {\em plateau estimate} in lattice determinations, and we will use this terminology here as well. The plateau estimate still depends on the source-sink separation due to excited states. Generically we find 
\begin{eqnarray}\label{simplnotationR}
g_{X,{\rm plat}}(t)\equiv R_X(t,t'=t/2) = g_X\Big[1+ {\rm O}\left(e^{-\Delta E_nt/2}\right) + {\rm O}\left( e^{-\Delta E_n t}\right) \Big]\,,
\end{eqnarray}
for the nucleon charges and, adjusting the prefactor, for the moments.

The main observation of the summation method \cite{Maiani:1987by,Capitani:2012gj} is that the ratio $R_X$ apparently has a stronger exponential suppression once the sum over all insertion times $t'$ is taken. More precisely, the {\em summation estimate} $s_X$ is defined as
\begin{equation}\label{defsumestimate}
s_X(t)\equiv \frac{d}{dt}\int_0^t dt' \, R_X(t,t')\,,
\end{equation}
and starting with  $R_X$ in \pref{DefRatio2} we obtain
\begin{equation}
s_X(t) = g_X + {\rm O}\left( e^{-\Delta E_n t}\right)\,.
\end{equation}
Compared with result \pref{simplnotationR} for the plateau estimate the exponential suppression is stronger. Still, in practice the summation estimate is not necessarily superior. The derivative typically results in larger statistical errors for the summation estimate: In lattice calculations one needs the integral in eq.\ \pref{defsumestimate} for at least three times $t$ and performs a linear fit to obtain the slope $g_X$. This typically results in larger statistical uncertainties compared with the plateau estimate.

\subsubsection{Plateau estimate}

\begin{figure}[t]
\begin{center}
$R_A(t,t/2)/g_A$\\
\includegraphics[scale=0.8]{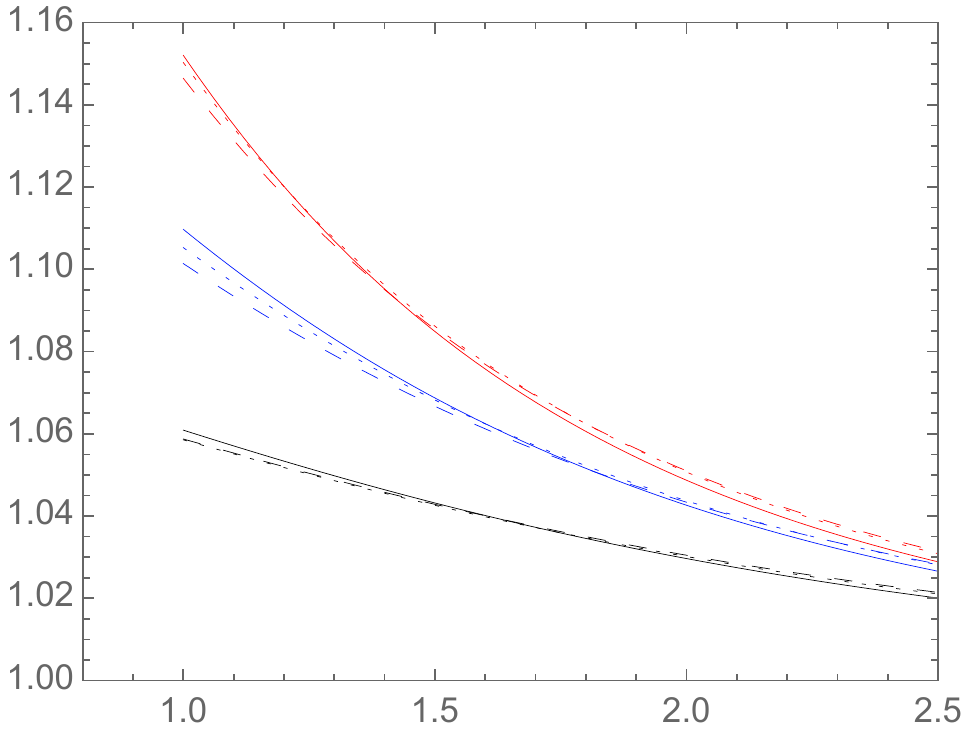}\\
$t$
\caption{The plateau estimate $R_A(t,t/2)$ normalized by $g_A$ as a function of the source-sink separation $t$. Results are shown for $M_{\pi}L=4$ (solid lines),  $M_{\pi}L=5$ (dotted lines) and $M_{\pi}L=6$ (dashed lines). Different colors distinguish between the three different energy bounds $E_{N\pi,{n_{\rm max}}}$ specified in table \ref{table:nmax}: first row in black, second row in blue and third row in red.}
\label{fig:axialcharge}
\end{center}
\end{figure}

Fig.\ \ref{fig:axialcharge} shows the ratio $R_A(t,t/2)/g_A$ as a function of the source-sink separation $t$. 
Without any excited-state contribution this ratio would be equal to 1. Any deviation from this constant value shows directly the $N\pi$ contribution to the plateau estimate in percent. Results are shown for the three different lattice sizes (solid, dashed and dotted lines)  and the three different energies $E_{N\pi,{n_{\rm max}}}$ (black, blue and red lines) specified in table \ref{table:nmax}. 

Qualitatively fig.\ \ref{fig:axialcharge} looks similar to fig. \ref{fig:effmass}. According to eq. \pref{DefRatio2} the ratio is equal to 1 plus a sum of exponentials. The ChPT results for the coefficients $b_{A,n}, \tilde{b}_{A,n}$ multiplying the dominant exponentials are positive, thus the ratio is larger than 1 and the $N\pi$ contribution in the plateau estimate leads to an overestimation of the axial charge. 

As for the effective mass the results for the three different $M_{\pi}L$ values lie essentially on top of each other; the differences are at the per mille level. We emphasize once again that this small FV dependence requires the energy interval of the $N\pi$ states taken into account kept fixed as $M_{\pi}L$ is varied.

A big difference compared to the effective mass results is that the source-sink separation needs to be significantly larger to suppress the high-momentum $N\pi$ states. To be specific let us consider again the $M_{\pi}L=4$ results. At $t=2.5$ fm the contribution of the first two $N\pi$ states (black curve) makes about 70 \% of the full contribution (red curve). To capture half of the full contribution $t$ still needs to be 1.5 fm. The blue curve, showing the $N\pi$ contribution of the lowest five $N\pi$ states, captures about 90\% for $t\approx 2$ fm. As discussed before, we want the low-momentum $N\pi$ states to dominate the excited-state contribution, since only for those we can expect a well-behaved chiral expansion with moderate higher-order corrections. Therefore, looking at fig. \ref{fig:axialcharge} we may conclude that source-sink separations of about 2 fm  
and larger are required for ChPT to make reliable estimates. How large the higher order corrections at these source-sink separations are is hard to predict. The naive error estimates we have made for the effective mass suggests here a 30-50\% uncertainty for $t\gtrsim2$ fm, and a more solid error estimate requires the calculation at NLO. With this caveat in mind we may conclude that the $N\pi$-state contribution in the plateau estimate leads to an overestimation of the axial charge by about 4\% at $t\approx 2$ fm.

\begin{figure}[t]
\begin{center}
$R_X(t,t/2)/g_X$ and $R_X(t,t/2)/\Pi_X$ \\
\includegraphics[scale=0.6]{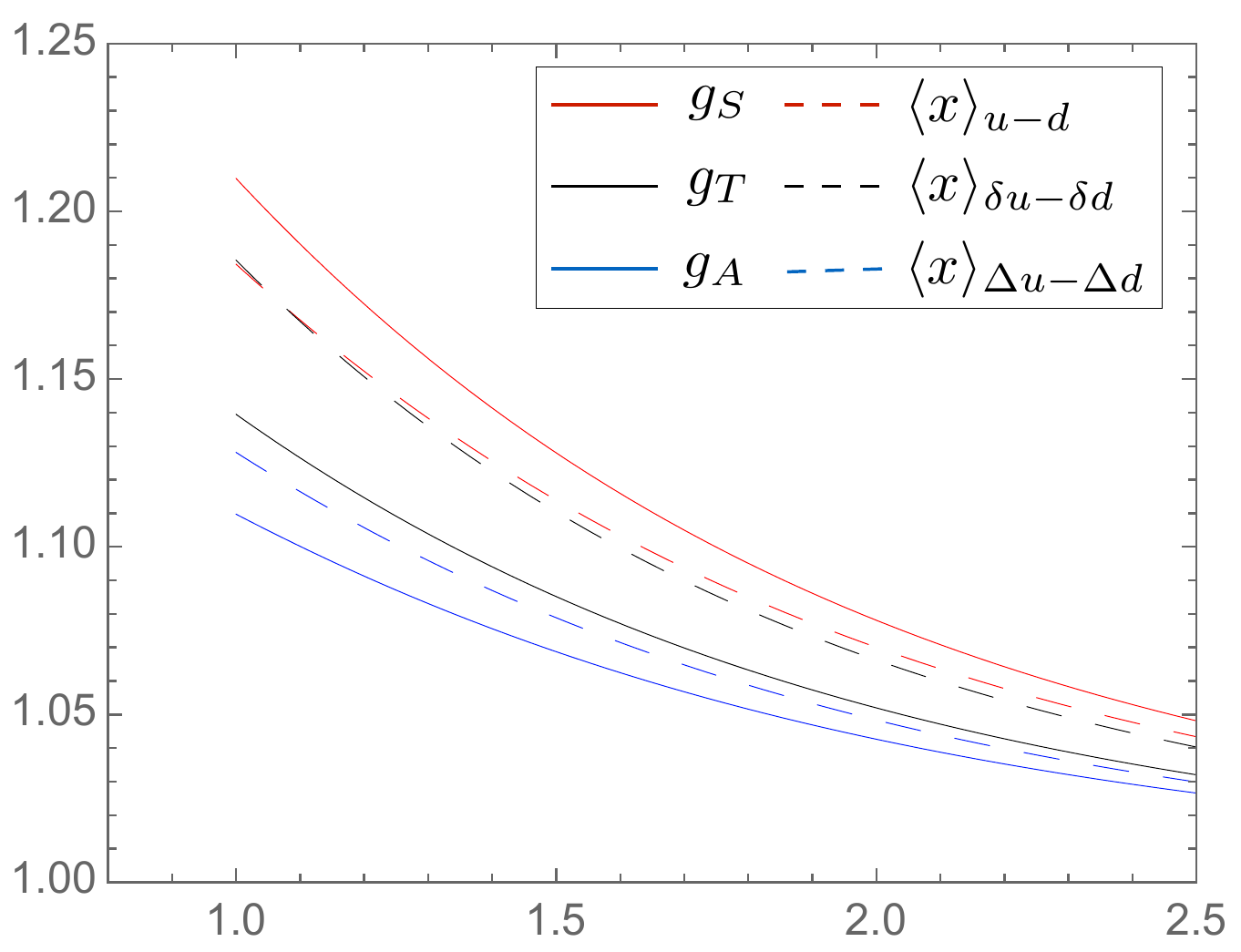}\\
$t$
\caption{The plateau estimates $R_X(t,t/2)$ normalized by the asymptotic values for all charges and moments. Results are shown for $M_{\pi}L=4$ and $p_{n_{\rm max}}/\Lambda_{\chi}=0.45$. The result for the axial charge (blue solid line) is the same as the blue solid line in fig.\ \ref{fig:axialcharge}.  }
\label{fig:resultsall}
\end{center}
\end{figure}

Qualitatively the same results are found for the tensor and scalar charges and the three moments. Fig.\ \ref{fig:resultsall} shows the plateau estimate normalized by either the charge or the moment. As for the axial charge we only find a tiny dependence on $M_{\pi}L$ in all cases, so fig.\ref{fig:resultsall}  shows the $M_{\pi}L=4$ results only. Moreover, only the results for $p_{n_{\rm max}}/\Lambda_{\chi}=0.45$ (second row in table \ref{table:nmax}) are displayed.\footnote{Figures showing the results for the other energy intervals can be found in Refs.\ \citen{Bar:2016uoj,Bar:2016jof}.} Although the source-sink separation in fig.\ \ref{fig:resultsall} starts at $t=1$ fm we find that, analogous to the axial charge, $t$ needs to be 2 fm or larger for the contribution of the high-momentum $N\pi$-states to be sufficiently suppressed. Therefore, the results displayed in fig.\ \ref{fig:resultsall} are expected to be reliable only for $t\gtrsim2$ fm.

Fig.\ \ref{fig:resultsall} shows that the $N\pi$-state contribution in the plateau estimates leads to an overestimation for all three charges and the three moments we consider here. The overestimation is smallest for the axial charge and largest for the scalar charge. For the latter the overestimation is about twice as big as for $g_A$. The $N\pi$ contribution to the average momentum fraction and the helicity moment is roughly of the same size as for the scalar charge. In summary we find an overestimation of about 5-10\% at $t\approx 2$ fm, and it slowly decreases to 3-6\% at $t\approx 2.5$ fm. 

\subsubsection{Comment on the summation estimate}
The summation method \cite{Maiani:1987by,Capitani:2012gj} starts from the ratio $R_X$ and computes the integral $S_X(t,t_{\rm m}) = \int^{t-t_{\rm m}}_{t_{\rm m}} dt' R_X(t,t')$.  As a function of $t$ (keeping $t_{\rm m}$ fixed) the slope is proportional to the moment one is interested in.
In actual lattice determinations $t_{\rm m}$ is taken to be essentially zero, so the integral is computed for all insertion times $t'$ between source and sink. On the other hand, for ChPT to give a good approximation of $R_X$ all time separations need to be large. Based on the results in the last section we need to require a minimal time separation of about 1 fm for $t_{\rm m}$ and $t-t_{\rm m}$. In addition we need a non-zero time interval $t-2t_{\rm m}$ to integrate over. This implies source-sink separations of at least 2.5 fm if not larger. Such large values are currently not accessible in lattice simulations, so here we do not consider the summation method any further. In case of the nucleon charges more details concerning the summation method can be found in Ref. \citen{Bar:2016uoj}.  

\subsection{Comparison with lattice data and discussion}

Some collaborations have already performed lattice calculations of the various charges and moments with a pion mass at or near the physical value \cite{Bali:2014nma,Abdel-Rehim:2015owa,Bhattacharya:2016zcn,vonHippel:2016wid,Green:2012ud,Bali:2014gha}. The main obstacle for directly applying the ChPT results to these lattice calculations are the fairly small source-sink separations in these simulations. In most cases the maximal source-sink separation $t_{\rm max}$ used to extract the charges and moments with the plateau method ranges between 1.1 and 1.3 fm. As discussed in the previous section, we do not expect ChPT to provide solid results unless the source-sink separations are about 2 fm or even larger. Still, it is useful to compare the existing lattice results with our ChPT results, since eventually the latter are expected to be reproduced.

Most of the existing lattice results for the charges and moments were obtained for pion masses larger than the physical value. There exist many reviews summarizing these results, and we simply refer to the most recent ones and the references therein.\cite{Syritsyn:2014saa,Green:2014vxa,Constantinou:2015agp,Alexandrou:2016hiy} Here we focus on results obtained with pion masses not larger than 150 MeV.
We are mainly interested in results for the axial charge, the average momentum fraction and the helicity moment. For these observables the experimental and phenomenological values are known rather precisely, thus we have data showing the discrepancy between these values and the lattice estimates.

In case of the axial charge Refs.\ \citen{Bali:2014nma,Abdel-Rehim:2015owa} provide plateau estimates as a function of the source-sink separation. In other references either the summation method \cite{Green:2012ud} or two-state fits \cite{Bhattacharya:2016zcn} were used to extract the axial charge, and these estimates cannot be compared with the ChPT results presented in the previous section.  

\begin{figure}[t]
\begin{center}
$g_{A,{\rm plat}}(t)/g_{A,{\rm exp}}$ \\
\includegraphics[scale=0.85]{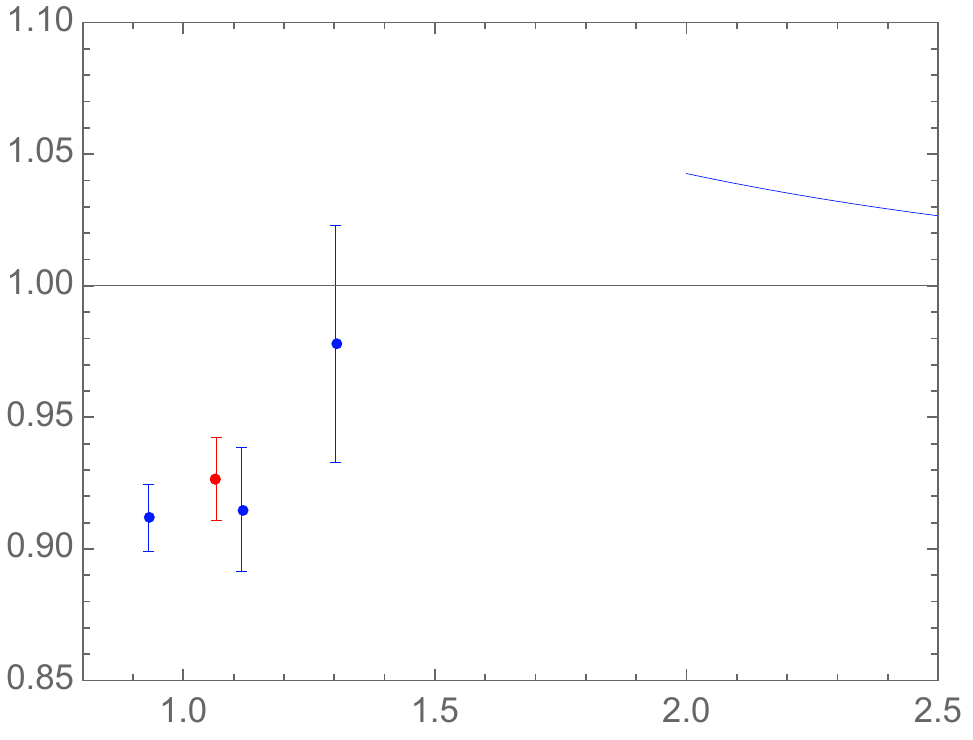}\\
$t$
\caption{Lattice plateau estimates for the axial charge normalized by the experimental value. Lattice data from Ref.\ \citen{Bali:2014nma} (red points) and Ref.\ \citen{Abdel-Rehim:2015owa} (blue points). The ChPT result of fig.\ \ref{fig:resultsall} is also shown for $t\ge 2$ fm by the blue solid line.}
\label{fig:compgA}
\end{center}
\end{figure}

Figure \ref{fig:compgA} shows the renormalized plateau estimates divided by the experimental value $g_{A,{\rm exp}} =1.2723(23)$\cite{Olive:2016xmw}, together with the ChPT result shown in fig.\ \ref{fig:resultsall}. The results of Ref.\ \citen{Abdel-Rehim:2015owa} (blue symbols) were obtained with $N_f=2$ twisted mass fermions with $M_{\pi}\approx 130$ MeV and $M_{\pi}L\approx 4.5$, while the remaining data point (red symbol), taken from Ref.\ \citen{Bali:2014nma}, was obtained with $N_f=2$ improved Wilson fermions with $M_{\pi} \approx 150$ MeV and $M_{\pi}L\approx 3.5$. The lattice spacings were $0.093$ fm and $0.071$ fm, respectively. More details about the simulation setup are given in the original references. 

The plateau estimates shown in fig.\ \ref{fig:compgA} were obtained for source-sink separations between 0.9 and 1.3 fm. The plateau estimates  are below the experimental value since $R_A(t,t/2)/g_{A,{\rm exp}} < 1$, although the estimate at $t=1.3$ fm agrees with $g_{A,{\rm exp}}$ given the rather large error bar. The ChPT prediction for the overestimation due to the $N\pi$ states is also shown, but we have truncated the result at $t=2$ fm. Apparently, there is still lots of room for the lattice data to connect smoothly with the ChPT prediction when $t$ is increased. 
In the most naive scenario $R_A(t,t/2)/g_{A,{\rm exp}}$ increases and crosses 1 at some $t$ between 1.5 fm and 2 fm, followed by a slow decrease to the asymptotic value. Whether this indeed happens needs to be checked, of course. This requires lattice calculations of $g_A$ at $t$ larger than 1.5 fm with a few percent precision.\footnote{In addition one should also check that the underestimation of the axial charge at small source-sink separations persists for smaller lattice spacings.} 

If indeed realized, such a scenario is potentially misleading in practice. Since the plateau estimate approaches the experimental value at some $t$ well before the asymptotic region is reached, one might be tempted to stop simulating at larger source-sink separations. In that case one reproduces the correct experimental value $g_{A,{\rm exp}}$, but for the wrong reason: The excited-state contributions are not small because $t$ is sufficiently large to be in the asymptotic regime, but various  excited-state contributions are non-vanishing and accidentally cancel each other. 

A concrete model for such a scenario was recently suggested by Hansen and Meyer.\cite{Hansen:2016qoz}. Based on plausible assumptions concerning the higher order corrections to the ChPT results, the high-momentum $N\pi$ states with energies larger than about 1.5 $M_N$ contribute {\em negatively}, i.e.\ the coefficients $b_n, \tilde{b}_n$ associated with these states are smaller than zero. Summing up the total $N\pi$ contamination the positive contribution from the low-momentum states is overcompensated by the contribution from the high-momentum states, leading to an underestimation of the axial charge for source-sink separations below $\sim 1.5$ fm, in agreement with lattice results.

\begin{figure}[t]
\begin{center}
$\langle x\rangle_{u-d,{\rm plat}}(t)/\langle x\rangle_{u-d,{\rm phen}}$ \\
\includegraphics[scale=0.85]{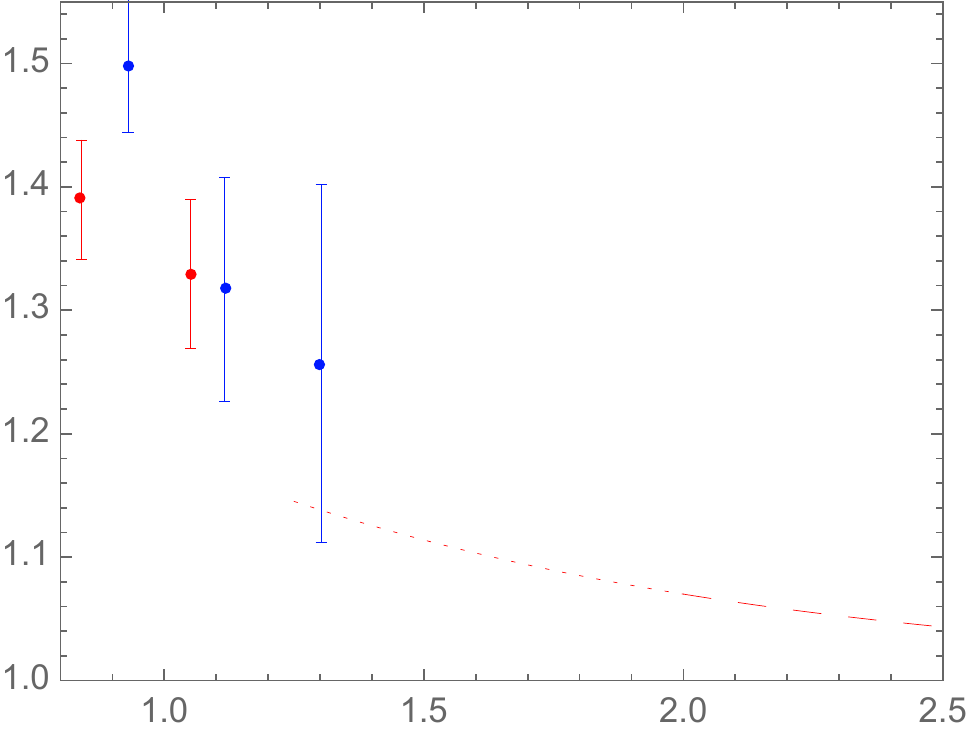}\\
$t$
\caption{Lattice plateau estimates for the average momentum fraction normalized by the phenomenological value. Phenomenological value from Ref.\ \citen{Alekhin:2012ig}, lattice data from Ref.\ \citen{Bali:2014gha} (red points) and Ref.\ \citen{Abdel-Rehim:2015owa} (blue points). The ChPT result of fig.\ \ref{fig:resultsall} is also shown for $t\ge 2$ fm by the red dashed line (dotted between $t=1.3$ fm and 2 fm where the ChPT result is expected to receive large higher order corrections).}
\label{fig:compavmom}
\end{center}
\end{figure}

Whether the high-momentum $N\pi$ states are indeed responsible for the compensation of the low-momentum ones needs to be corroborated. Other excited states (e.g.\ $N\pi\pi$ and $\Delta\pi$ states) may also play a non-negligible role. In any case, the model in Ref.\ \citen{Hansen:2016qoz} supports our expectation that source-sink separations $t\gtrsim2$ fm are needed for the LO ChPT results to be applicable. 

\begin{figure}[t]
\begin{center}
$\langle x\rangle_{\Delta u-\Delta d,{\rm plat}}(t)/\langle x\rangle_{\Delta u-\Delta d,{\rm phen}}$ \\
\includegraphics[scale=0.85]{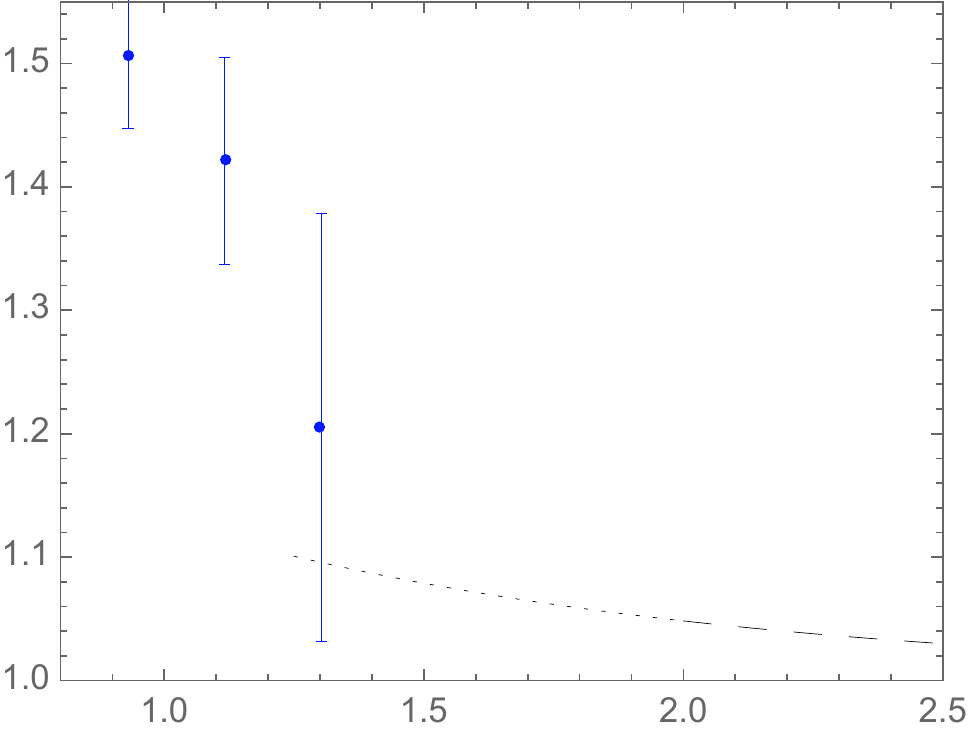}\\
$t$
\caption{Lattice plateau estimates for the helicity moment normalized by the phenomenological value. Phenomenological value from Ref.\ \citen{Blumlein:2010rn}, lattice data from Ref.\ \citen{Abdel-Rehim:2015owa} (blue points). The ChPT result of fig.\ \ref{fig:resultsall} is also shown for $t\ge 2$ fm by the black dashed line (dotted between $t=1.3$ fm and 2 fm where the ChPT result is expected to receive large higher order corrections).}
\label{fig:comphelmom}
\end{center}
\end{figure}

Figs.\ \ref{fig:compavmom} and \ref{fig:comphelmom} show the plateau estimates given in Refs.\ \citen{Abdel-Rehim:2015owa,Bali:2014gha} for $\mf$ and $\hm$, normalized by their phenomenological values. The phenomenological values are (in the $\overline{\rm MS}$ scheme at 2 GeV) $\langle x\rangle_{u-d,{\rm phen}}=0.1655(39)$ and $\langle x\rangle_{\Delta u-\Delta d,{\rm phen}}=0.190(8)$\cite{Alekhin:2012ig,Blumlein:2010rn}. 
The ensemble used in Ref.\ \citen{Bali:2014gha} is the same as the one used in Ref.\ \citen{Bali:2014nma} to compute the axial charge (corresponding to the red dots in fig.\ \ref{fig:compgA}).  
 
For both observables the plateau estimates overestimate the phenomenological value. The absolute discrepancy is significantly larger than for the axial charge (note the different scales in figs.\ \ref{fig:compavmom} and \ref{fig:comphelmom}).  The statistical errors are quite large and increase with $t$ getting larger. Still, the plots suggest that the plateau estimates decrease as the source-sink separation increases.  Here too the lattice data and the ChPT predictions for the ratios are consistent. Compared with the results for the axial charge it is here much simpler to imagine a simple monotonic decrease until contact with ChPT can be made at source sink-separations between 2 and 2.5 fm. 

It might be tempting to subtract the $N\pi$ contamination predicted by ChPT (dotted lines) to achieve agreement between the lattice results and the phenomenological values. However, since the same subtraction fails dismally in case of the axial charge it should neither be attempted for $\mf$ and $\hm$.  
 
In case of the scalar and tensor charges  as well as the transversity moment we do not have experimental values at our disposal. Thus, comparisons between the lattice plateau estimates and the ChPT predictions cannot be done. However, there is no reason to believe that the asymptotic region is reached at significantly smaller source-sink separations for these observables. In fact, the absence of experimental results  
elevates the role ChPT can play as a guide for the systematic uncertainties due excited states, provided one is in the asymptotic regime where ChPT can be applied.
 
The bottom line of this section is that present lattice data at source-sink separations $t\lesssim 1.3$ fm is still far away from the asymptotic regime where the lowest lying $N\pi$ states are responsible for the dominant excited-state contribution in the plateau estimates. Lattice data at significantly larger source-sink separations (and with sufficiently small statistical errors) are needed to make contact with the asymptotic region where ChPT can be used to extrapolate the data to $t\rightarrow \infty$.
 
\section{Concluding remarks}

Our findings can be summarized as follows. The $N\pi$ state contribution in the plateau estimates leads to an overestimation of all six nucleon observables we considered, the axial, scalar and tensor charges and the average momentum fraction, the helicity and the transversity moment. 
The overestimation is at the 5-10\% level assuming source-sink separations of about 2 fm. These results are based on LO ChPT and may be subject of substantial higher order corrections. Still, the $N\pi$ contamination is uncomfortably large and certainly not ignorable if results with percent precision are the goal of lattice calculations. 

In physical point simulations the $N\pi$-state contribution is expected to dominate the excited-state contamination, at least in the asymptotic regime with large but still finite source-sink separations. The smaller $t$ the larger are the higher order corrections to the LO results presented here. A calculation of the NLO corrections is certainly desirable in order to obtain firmer error estimates for the LO results.
In addition, it needs to be checked that the contributions of three-particle $N\pi\pi$-states and two-particle $\Delta \pi$ states are small and under control. ChPT calculations of these contributions are essentially analogous to the calculation of the $N\pi$ contribution.

Even though the calculation of the higher-order corrections and the additional multi-particle contribution is in principle straightforward one should keep in mind that the results of these calculations will involve some additional LECs that are not known and not easily accessible using experimental input. Examples are the LECs associated with the interpolating nucleon fields or the coupling of the nucleon and the delta. How important these contributions are remains to be seen. However, the goal cannot be to determine the unknown LECs by fits to lattice data. Such fits seem to be doomed from the beginning, since these parameters appear in higher order corrections of a O(10\%) excited-state contamination in a hadronic correlation function that is prone to large statistical errors. It seems more promising to make conservative estimates concerning the unknown coefficients and vary them within generously chosen bounds. Based  on these values one should be able to estimate the source-sink separations that are necessary for the LO results to apply reasonably well, and the higher order corrections and the additional contributions can be included in a systematic error for the LO result.

Irrespective of how large the final error estimates for the LO results will be, source-sink separations as small as they are in contemporary lattice simulations are far too small for ChPT to be applicable. In order to make contact to the ChPT results lattice data at significantly larger $t$ with significantly smaller statistical errors are needed. This will require new simulation techniques to overcome the signal-to-noise problem in lattice calculations. New ideas in that respect have been recently been proposed\cite{Ce:2016idq,Ce:2016ajy,Wagman:2017xfh}, but remain to be tested in actual lattice calculations of the nucleon correlation functions considered here.

\vspace{2ex}
\noindent {\bf Acknowledgments}
\vspace{2ex}

I thank Rainer Sommer for comments on a first draft of this manuscript. 
This work was supported by the Japan Society for the Promotion of Science (JSPS) with an Invitation Fellowship for Research in Japan (ID No. L16520), and by the German Research Foundation (DFG), Grant ID BA 3494/2-1.


\begin{thebibliography}{10}

\bibitem{Bar:2012ce}
O.~B{\"a}r and M.~Golterman,
\newblock Phys. Rev. {\bf D87} (2013) 014505.

\bibitem{Tiburzi:2009zp}
B.~C. Tiburzi,
\newblock Phys. Rev. {\bf D80} (2009) 014002.

\bibitem{Bar:2015zwa}
O.~B{\"a}r,
\newblock Phys. Rev. {\bf D92} (2015) 074504.

\bibitem{Tiburzi:2015tta}
B.~C. Tiburzi,
\newblock Phys. Rev. {\bf D91} (2015) 094510.

\bibitem{Bar:2016uoj}
O.~B{\"a}r,
\newblock Phys. Rev. {\bf D94} (2016) 054505.

\bibitem{Bar:2016jof}
O.~B{\"a}r,
\newblock Phys. Rev. {\bf D95} (2017) 034506.

\bibitem{Luscher:2013vga}
M.~L{\"u}scher,
\newblock PoS {\bf LATTICE2013} (2014) 016.

\bibitem{Bar:2013ora}
O.~B{\"a}r and M.~Golterman,
\newblock Phys.Rev. {\bf D89} (2014) 034505.

\bibitem{Bar:BetheForum}
{O.\ B{\"a}r},
\newblock {{\em Nucleon-pion-state contributions to nucleon correlation
  functions}, talk given at the workshop {\em Methods in Lattice Field theory},
  Bethe Forum, Bonn, Germany, 2015.}

\bibitem{Bar:2015zha}
O.~B{\"a}r,
\newblock {{\em Nucleon-pion-state contributions in the determination of the nucleon
  axial charge}},
\newblock in {\em {Proceedings, 33rd International Symposium on Lattice Field
  Theory (Lattice 2015)}}, 2015.

\bibitem{Bar:Niigata}
{O.\ B{\"a}r},
\newblock { {\em $N\pi$-state contribution in lattice calculations of the
  nucleon charges $g_A,g_T$ and $g_S$}, talk given at the workshop {\em Phase
  structure of lattice field theories}, Japanese-German-Seminar, Niigata,
  Japan, 2016.}

\bibitem{Parisi:1983ae}
G.~Parisi,
\newblock Phys. Rept. {\bf 103} (1984) 203.

\bibitem{Lepage:1989hd}
G.~P. Lepage,
\newblock {The Analysis of Algorithms for Lattice Field Theory},
\newblock in {\em {Boulder ASI 1989:97-120}}, pp. 97--120, 1989.

\bibitem{Luscher:1991cf}
M.~L{\"u}scher,
\newblock Nucl. Phys. {\bf B364} (1991) 237.

\bibitem{Yoon:2016jzj}
B.~Yoon {\em et~al.},
\newblock arXiv:1611.07452[hep-lat].

\bibitem{Luscher:1990ck}
M.~L{\"u}scher and U.~Wolff,
\newblock Nucl.Phys. {\bf B339} (1990) 222.

\bibitem{Lang:2012db}
C.~Lang and V.~Verduci,
\newblock Phys.Rev. {\bf D87} (2013) 054502.

\bibitem{Kiratidis:2015vpa}
A.~L. Kiratidis, W.~Kamleh, D.~B. Leinweber and B.~J. Owen,
\newblock Phys. Rev. {\bf D91} (2015) 094509.

\bibitem{Lang:2016hnn}
C.~B. Lang, L.~Leskovec, M.~Padmanath and S.~Prelovsek,
\newblock Phys. Rev. {\bf D95} (2017) 014510.

\bibitem{Kiratidis:2016hda}
A.~L. Kiratidis {\em et~al.},
\newblock Phys. Rev. {\bf D95} (2017) 074507.

\bibitem{Olive:2016xmw}
C.~Patrignani {\em et~al.},
\newblock Chin. Phys. {\bf C40} (2016) 100001.

\bibitem{Bhattacharya:2011qm}
T.~Bhattacharya {\em et~al.},
\newblock Phys. Rev. {\bf D85} (2012) 054512.

\bibitem{Abdel-Rehim:2015owa}
A.~Abdel-Rehim {\em et~al.},
\newblock Phys. Rev. {\bf D92} (2015) 114513,
\newblock [Erratum: Phys. Rev.D93,no.3,039904(2016)].

\bibitem{Weinberg:1978kz}
S.~Weinberg,
\newblock Physica {\bf A96} (1979) 327.

\bibitem{Gasser:1983yg}
J.~Gasser and H.~Leutwyler,
\newblock Ann. Phys. {\bf 158} (1984) 142.

\bibitem{Gasser:1984gg}
J.~Gasser and H.~Leutwyler,
\newblock Nucl. Phys. {\bf B250} (1985) 465.

\bibitem{Colangelo:2000zw}
G.~Colangelo and G.~Isidori,
\newblock {An Introduction to ChPT},
\newblock in {\em {Nuclear, subnuclear and astroparticle physics. Proceedings,
  5th LNF Spring School, Frascati, Italy, May 15-20, 2000}}, pp. 333--376,
  2000.

\bibitem{Gasser:2003cg}
J.~Gasser,
\newblock Lect. Notes Phys. {\bf 629} (2004) 1.

\bibitem{Golterman:2009kw}
M.~Golterman,
\newblock {Applications of chiral perturbation theory to lattice QCD},
\newblock in {\em {Modern perspectives in lattice QCD: Quantum field theory and
  high performance computing. Proceedings, International School, 93rd Session,
  Les Houches, France, August 3-28, 2009}}, pp. 423--515, 2009.

\bibitem{Scherer:2012xha}
S.~Scherer and M.~R. Schindler,
\newblock Lect.Notes Phys. {\bf 830} (2012) pp.1.

\bibitem{Ecker:2013xja}
G.~Ecker,
\newblock Nucl. Phys. Proc. Suppl. {\bf 245} (2013) 1.

\bibitem{Gasser:1987rb}
J.~Gasser, M.~Sainio and A.~Svarc,
\newblock Nucl.Phys. {\bf B307} (1988) 779.

\bibitem{Becher:1999he}
T.~Becher and H.~Leutwyler,
\newblock Eur.Phys.J. {\bf C9} (1999) 643.

\bibitem{Bernard:2007zu}
V.~Bernard,
\newblock Prog. Part. Nucl. Phys. {\bf 60} (2008) 82.

\bibitem{Kubis:2007iy}
B.~Kubis,
\newblock {An Introduction to chiral perturbation theory},
\newblock in {\em {Workshop on Physics and Astrophysics of Hadrons and Hadronic
  Matter Shantiniketan, India, November 6-10, 2006}}, 2007.

\bibitem{Scherer:2009bt}
S.~Scherer,
\newblock Prog. Part. Nucl. Phys. {\bf 64} (2010) 1.

\bibitem{Gasser:1983ky}
J.~Gasser and H.~Leutwyler,
\newblock Phys.Lett. {\bf B125} (1983) 321.

\bibitem{Fettes:2000gb}
N.~Fettes, U.-G. Meissner, M.~Mojzis and S.~Steininger,
\newblock Annals Phys. {\bf 283} (2000) 273.

\bibitem{Georgi:1990um}
H.~Georgi,
\newblock Phys. Lett. {\bf B240} (1990) 447.

\bibitem{Jenkins:1990jv}
E.~Jenkins and A.~V. Manohar,
\newblock Phys. Lett. {\bf B255} (1991) 558.

\bibitem{Dorati:2007bk}
M.~Dorati, T.~A. Gail and T.~R. Hemmert,
\newblock Nucl. Phys. {\bf A798} (2008) 96.

\bibitem{Wein:2014wma}
P.~Wein, P.~C. Bruns and A.~Sch{\"a}fer,
\newblock Phys. Rev. {\bf D89} (2014) 116002.

\bibitem{Cata:2007ns}
O.~Cata and V.~Mateu,
\newblock JHEP {\bf 09} (2007) 078.

\bibitem{Ioffe:1981kw}
B.~Ioffe,
\newblock Nucl.Phys. {\bf B188} (1981) 317.

\bibitem{Espriu:1983hu}
D.~Espriu, P.~Pascual and R.~Tarrach,
\newblock Nucl.Phys. {\bf B214} (1983) 285.

\bibitem{Nagata:2008zzc}
K.~Nagata, A.~Hosaka and V.~Dmitrasinovic,
\newblock Eur.Phys.J. {\bf C57} (2008) 557.

\bibitem{Wein:2011ix}
P.~Wein, P.~C. Bruns, T.~R. Hemmert and A.~Sch{\"a}fer,
\newblock Eur.Phys.J. {\bf A47} (2011) 149.

\bibitem{Gusken:1989ad}
S.~G{\"u}sken {\em et~al.},
\newblock Phys.Lett. {\bf B227} (1989) 266.

\bibitem{Gusken:1989qx}
S.~G{\"u}sken,
\newblock Nucl.Phys.Proc.Suppl. {\bf 17} (1990) 361.

\bibitem{Alexandrou:1990dq}
C.~Alexandrou {\em et~al.},
\newblock Phys.Lett. {\bf B256} (1991) 60.

\bibitem{Luscher:2013cpa}
M.~L{\"u}scher,
\newblock JHEP {\bf 1304} (2013) 123.

\bibitem{Hansen:2016qoz}
M.~T. Hansen and H.~B. Meyer,
\newblock arXiv:1610.03843[hep-lat].

\bibitem{Lellouch:2000pv}
L.~Lellouch and M.~L{\"u}scher,
\newblock Commun. Math. Phys. {\bf 219} (2001) 31.

\bibitem{Alekhin:2012ig}
S.~Alekhin, J.~Bl{\"u}mlein and S.~Moch,
\newblock Phys. Rev. {\bf D86} (2012) 054009.

\bibitem{Blumlein:2010rn}
J.~Bl{\"u}mlein and H.~B{\"o}ttcher,
\newblock Nucl. Phys. {\bf B841} (2010) 205.

\bibitem{Colangelo:2003hf}
G.~Colangelo and S.~D{\"u}rr,
\newblock Eur.Phys.J. {\bf C33} (2004) 543.

\bibitem{Ishikawa:2015rho}
K.~I. Ishikawa {\em et~al.},
\newblock PoS {\bf LATTICE2015} (2016) 075.

\bibitem{Bali:2014nma}
G.~S. Bali {\em et~al.},
\newblock Phys. Rev. {\bf D91} (2015) 054501.

\bibitem{Maiani:1987by}
L.~Maiani, G.~Martinelli, M.~L. Paciello and B.~Taglienti,
\newblock Nucl. Phys. {\bf B293} (1987) 420.

\bibitem{Capitani:2012gj}
S.~Capitani {\em et~al.},
\newblock Phys. Rev. {\bf D86} (2012) 074502.

\bibitem{Bhattacharya:2016zcn}
T.~Bhattacharya {\em et~al.},
\newblock arXiv:1606.07049 [hep-lat].

\bibitem{vonHippel:2016wid}
G.~von Hippel, T.~D. Rae, E.~Shintani and H.~Wittig,
\newblock Nucl. Phys. {\bf B914} (2017) 138.

\bibitem{Green:2012ud}
J.~R. Green {\em et~al.},
\newblock Phys. Lett. {\bf B734} (2014) 290.

\bibitem{Bali:2014gha}
G.~S. Bali {\em et~al.},
\newblock Phys. Rev. {\bf D90} (2014) 074510.

\bibitem{Syritsyn:2014saa}
S.~Syritsyn,
\newblock PoS {\bf LATTICE2013} (2014) 009.

\bibitem{Green:2014vxa}
J.~Green,
\newblock AIP Conf. Proc. {\bf 1701} (2016) 040007.

\bibitem{Constantinou:2015agp}
M.~Constantinou,
\newblock PoS {\bf CD15} (2015) 009.

\bibitem{Alexandrou:2016hiy}
C.~Alexandrou,
\newblock {Novel applications of Lattice QCD: Parton distribution functions,
  proton charge radius and neutron electric dipole moment},
\newblock arXiv:1612.04644[hep-lat].

\bibitem{Ce:2016idq}
M.~C\`{e}, L.~Giusti and S.~Schaefer,
\newblock Phys. Rev. {\bf D93} (2016) 094507.

\bibitem{Ce:2016ajy}
M.~C\`{e}, L.~Giusti and S.~Schaefer,
\newblock Phys. Rev. {\bf D95} (2017) 034503.

\bibitem{Wagman:2017xfh}
M.~L. Wagman and M.~J. Savage,
\newblock arXiv:1704.07356 [hep-lat].

\end{thebibliography}
\end{document}